%% file: main.tex
\documentclass[amssymb,amsmath,prc,twocolumn,preprintnumbers,superscriptaddress,nofootinbib, tik]{revtex4-2}

\usepackage{graphicx}
\usepackage[usenames,dvipsnames]{xcolor}
\usepackage{tikz}
\usetikzlibrary{arrows.meta, positioning, shapes.geometric, fit, backgrounds}

\usepackage{bm}
\usepackage{amssymb, amsmath, amsfonts}
\usepackage{mathrsfs}
\usepackage{latexsym}
\usepackage{color}
\usepackage[normalem]{ulem} 
\usepackage{dcolumn}

\usepackage{xspace}
\usepackage{graphicx}
\usepackage{booktabs}
\usepackage{siunitx}
\usepackage{nccmath}
\usepackage{dsfont}
\usepackage{multirow}
\usepackage{xfp} 
\usepackage{comment}
\usepackage{nameref}
\usepackage{orcidlink}
\usepackage[capitalise]{cleveref}
\hypersetup{
citecolor=blue,
 colorlinks=true,
 linkcolor=blue,
 filecolor=magenta,
 urlcolor=blue,
}
\crefname{section}{Sec.}{Secs.}
\crefname{equation}{Eq.}{Eqs.}
\crefname{figure}{Fig.}{Figs.}
\crefname{appendix}{App.}{Apps.}

\newcommand{\mev}{\mathrm{MeV}}
\DeclareMathOperator{\SNM}{SNM}
\DeclareMathOperator{\PNM}{PNM}
\newcommand{\sat}{\mathrm{sat}}
\newcommand{\Esat}{E_{B}}
\newcommand{\nsat}{n_{\sat}}
\newcommand{\Esym}{J}
\newcommand{\Lsym}{L}

\newcommand{\Ksat}{K_{0}}

\newcommand{\suthree}[1]{\mathrm{SU(3)}_{\mathrm{#1}}}
\newcommand{\Tr}[1]{\mathrm{Tr}\left(#1\right)} 
\newcommand{\Ccoupling}{singlet-octet }

\newcommand{\physcoupling}{physical-meson}

\newcommand{\cmfpp}{{\sc CMF++} }
\newcommand{\cmf}{CMF}

\newcommand{\red}[1]{\textcolor{red}{#1}}

\newcommand{\Z}{\mathbb{Z}}

\newcommand{\half}{\frac{1}{2}}

\usepackage[english]{babel}
\usepackage{xspace}
\usepackage{color,soul}

\newcommand{\result}[1]{\red{#1}}
\newcommand{\quant}[1]{\left(#1\right)}

\newcommand{\UIUC}{\affiliation{Illinois Center for Advanced Studies of the Universe, Department of Physics, University of Illinois at Urbana-Champaign, \\Urbana, Illinois, 61801, USA}}

\begin{document}
\include{macros}

\title{Neural-Accelerated Bayesian Calibration of Chiral Mean-Field Models to Nuclear Saturation and Vacuum Properties}

\author{Isaac Legred}\email{ilegred2@illinois.edu} \UIUC
\author{Mateus Reinke Pelicer \orcidlink{0000-0002-2189-706X}} \email{matrp@illinois.edu} \UIUC
\author{Veronica Dexheimer} \affiliation{Center for Nuclear Research, Department of Physics, Kent State University, Kent, OH 44243, USA}
\author{Jacquelyn Noronha-Hostler \orcidlink{0000-0003-3229-4958}}  \UIUC
\author{Nicol\'as Yunes \orcidlink{0000-0001-6147-1736}} \UIUC
\begin{abstract}

Chiral models of nuclear interactions provide approximate, phenomenological descriptions of dense matter that respect the symmetries of quantum chromodynamics. 
Their Lagrangian parameters, however, are difficult to calibrate because these models are not controlled effective theories.  Furthermore, repeated model evaluations are computationally expensive, and most parameter choices fail to reproduce acceptable saturation properties or hadron masses in vacuum.
To address this, we develop a Bayesian inference framework to identify parameter regions consistent with nuclear saturation properties and vacuum experimental constraints. 
We implement this framework through a neural-network surrogate approximation that accelerates the repeated mapping from model parameters to nuclear and particle observables.
Our fully-modular, neural-accelerated Bayesian framework interfaces the open-source MUSES Calculation Engine, the bilby inference library, and the PyTorch machine-learning toolkit.
We then apply the framework to the chiral mean-field model with a new generalized quartic vector self-interaction sector. 
We find that viable solutions are rare but broadly distributed within certain regions of parameter space, with the data constraining combinations of couplings more strongly than individual Lagrangian parameters.  
The resulting degeneracies imply that distinct saturation-compatible models can lead to qualitatively different descriptions of dense nuclear matter and, thus, of neutron stars, highlighting the need to combine terrestrial and astrophysical information. 

\end{abstract}

\maketitle

\section{Introduction}

The equation of state (EoS) of cold, beta-equilibrated dense matter in neutron stars (NSs) is connected to observations of nearly symmetric matter in much less extreme conditions on Earth \cite{Lattimer:2012nd,Oertel:2016bki,Drischler:2021kxf}.  The symmetries and approximate symmetries of quantum chromodynamics (QCD) restrict the possible functional forms of effective Lagrangians in the regime that is relevant for dense matter \cite{Weinberg:1990rz}. Therefore, within a particular phenomenological model, it is possible to place relatively tight constraints on the EoS of nuclear matter inside NSs, using observations of symmetric matter in terrestrial experiments \cite{Lattimer:2012nd, Kortelainen:2010hv, Moller:2012pxr, Oyamatsu:2022twl, Danielewicz:2002pu, Reed:2021nqk}.  However, this identification carries substantial systematic uncertainty: without a clear connection to QCD, one cannot guarantee that the model contains all of the relevant particles and interactions in the regime where it is applied.  Nonetheless, using internally-consistent microphysical models for the dense matter EoS is crucial for quantifying the \emph{dynamics} of NSs, including out-of-equilibrium effects \cite{Sawyer:1989dp, Espino:2023dei, HegadeKR:2026iou}.

One common class of dense-matter models use relativistic mean-field (RMF) theory to compute the equation of state~\cite{Walecka1974,Serot:1984ey,Ring:1996qi}.  Such models are constructed by treating dense nuclear matter as a collection of relativistic fermions (hadrons and quarks) interacting via the exchange of mesons.  Mesons in this framework are subsequently treated in the \emph{mean-field} approximation, indicating occupation numbers are sufficiently large that fields can be replaced by their expectation values.  This approximation renders such models tractable, and these models also are capable of reproducing nuclear matter observables with reasonable accuracy~\cite{Ring:1996qi,Dutra:2014qga}.   

A particular model of nuclear matter is the chiral mean-field (\cmf) model, which allows the inclusion of nucleons, hyperons, decuplet baryons and quark degrees of freedom self-consistently in the mean-field approximation~\cite{Papazoglou:1998vr,Dexheimer:2009hi}.  While the model is not derived explicitly as a low-energy effective theory from QCD, it is built from the approximate chiral symmetry of QCD and explicit symmetry breaking in the non-linear realization of $\suthree{}$. We use the open-source \cmfpp software~\cite{zenodo_cmf} for efficient evaluations of the EoS. The \cmf{} model has been shown to faithfully reproduce the properties of symmetric nuclear matter and closed-shell nuclei, and it is consistent with the existence of massive $(2M_{\odot})$ NSs \cite{Papazoglou:1998vr,Kumar:2024owe}.  However, it is normally challenging for \cmf{} and RMF models in general to reach ultraheavy neutron stars (e.g. $2.3-2.5 M_{\odot}$), which may require extra fine tuning, new features and/or considering rotation \cite{Dexheimer:2020rlp,Kumar:2023mlp}.

Compared with Walecka-like RMF models, the \cmf{} parameter space is harder to explore systematically. In many RMF models, baryon-meson and meson self-interactions are introduced as relatively independent phenomenological terms, making it possible to tune particular interaction channels directly. This allows constraints to be placed on parameters directly~\cite{Traversi:2020aaa,Malik:2023mnx,Scurto:2024ekq,Xie:2026hwg,Ma:2026dta}, as correlations between physical effects are limited.   In chiral models, by contrast, the underlying flavor structure correlates the allowed interactions. In the \cmf{} model, baryon-meson couplings are built from singlet and octet structures, while meson self-interactions must arise from chiral-invariant combinations. As a result, changing one Lagrangian parameter can modify several physical interaction channels at once.
This feature of the \cmf{} model makes Bayesian calibration of the model qualitatively different from calibration in simpler RMF parametrizations. Previous \cmf{} studies have used selected combinations of chiral vector interactions~\cite{Dexheimer:2015qha}, added non-chiral terms to reproduce nuclear and astrophysical data~\cite{Dexheimer:2018dhb,Dexheimer:2020rlp}, or performed Bayesian inference in a reduced nucleonic sector~\cite{Malik:2024qjw}. 

Here, we instead retain the broader quartic invariant basis relevant to the full \cmf{} flavor structure. In particular,  our analysis uses an extended set of quartic vector self-interactions allowed by chiral symmetry.  This interaction framework reduces to various interaction schemes used in the literature as limiting cases~\cite{Dexheimer:2015qha,Malik:2024qjw}. 

We show explicitly how the flavor structure of the model constrains the vector self-interaction couplings by rewriting them in the diagonal singlet-octet basis and in the operator basis. These terms are not all independently constrainable, so the set of acceptable values of each is neither obvious nor easily computable.

We then use a hierarchical Bayesian inference scheme to identify regions of \cmf{} parameter space that are consistent with nuclear matter near saturation. The hierarchy is useful because different observables probe different parts of the model. In the \cmf{} vacuum, the nucleon and hyperon masses can be computed analytically from the scalar-sector couplings. We can therefore constrain the vacuum parameters first, and then use the resulting posterior as input for the in-medium inference.

The in-medium problem is more expensive. Although the mean-field approximation makes each EoS calculation tractable, the saturation properties must still be obtained by solving the coupled mean field equations of motion for each choice of model parameters. A stochastic exploration of the corresponding high-dimensional posterior can require millions of likelihood evaluations. Thus, even if one EoS evaluation takes only $\mathcal O(10)$ seconds, a direct inference can require of order a year of CPU time. We therefore use neural-network emulators to accelerate the map from \cmfpp parameters to vacuum and saturation observables. We find that relatively simple networks reproduce this map with sufficient fidelity for exploratory Bayesian inference.

Neural networks are function approximation schemes that use alternating matrix multiplication and very simple nonlinear functions to approximate arbitrarily complex behavior.  The key insight (which we make use of in \Cref{sec:nuclear-matter-results}) is that such a function is very straightforward to differentiate, which streamlines optimization of the network.  Substantial developments in the field of machine learning have rendered such tools easy to use and highly computationally efficient.  Formally speaking, even the error induced by using a model approximation can be mitigated using \emph{likelihood reweighting}, sometimes called \emph{importance sampling}~\cite{Kloek:1978}, which reevaluates the exact likelihood on the relevant samples in the posterior.  

We find that the nuclear matter constraints act primarily on combinations of parameters, rather than on individual parameters.  This is in agreement with existing work studying the local structure of observables in \cmfpp~\cite{Cruz-Camacho:2026pdg}.  Further, though, we find that while the vast majority ($\gg 99\%$ of uniformly sampled configurations) of possible parameter combinations are excluded by nuclear matter constraints, the posterior distribution of parameters remains broad. This is characteristic of relatively low-dimensional structures with high likelihood inhabiting broadly the full higher-dimensional space of the prior.  These high-likelihood points are not localized, because degeneracies allow very different parameter combinations to produce the same physical observables.  Thus, correlations between parameters are crucial in the structure of the posterior.    In addition, the  diverse regions of parameter space these points occupy may produce vastly different astrophysical observables (reaching up to even maximum mass of $2.3 M_{\odot}$ for nucleonic equations of state). This highlights the important synergies between terrestrial and astrophysical constraints in determining the properties of dense matter.
The paper is organized as follows: In \Cref{sec:cmf-descripton} we give a brief overview of the \cmf{} model, and in \Cref{sec:cmf-couplings} we discuss the couplings that will be explored in this work and their relation to vacuum and in-medium observables. In \Cref{sec:bayesian_analysis}, we discuss our strategy for using Bayesian inference to explore the parameter space of the \cmf{} model.   In \Cref{sec:bayesian-theory}, we 
describe our strategy for sequentially incorporating information from vacuum and in-medium probes of nuclear physics respectively.  In \Cref{sec:emulator-and-iterative-scheme}, we discuss our strategy for using neural-network emulators of the mappings from parameters to observables.
In \Cref{sec:results} we discuss our results.  We first identify a set of parameters in the \cmfpp model that can be fixed solely from measurements of particle masses in vacuum.  Because the masses we use are measured very precisely, we demonstrate that systematic uncertainty dominates constraints on these parameters.  We go on to discuss constraints on parameters that have an effect only in-medium by considering observables defined at or near nuclear saturation density ($\nsat$).  Here we identify  largely unexplored regions of parameter space that are fully consistent with constraints on nuclear matter, but allow for a wide variety of NS properties.   We present conclusions in \Cref{sec:discussion}.

\section{The Chiral Mean Field Model}
\label{sec:cmf-descripton}

The chiral mean field model is a relativistic framework for baryon and quark interactions in terms of classical scalar and vector meson background fields, with the allowed interactions constrained by the underlying $\suthree{L}\times\suthree{R}$ flavor structure~\cite{Papazoglou:1997uw,Papazoglou:1998vr,Dexheimer:2008ax,Dexheimer:2009hi}. It is based on the non-linear realization of chiral symmetry.
The scalar fields generate the attractive part of the interaction by modifying the in-medium fermion masses, while the vector fields generate the repulsive interactions by shifting the effective chemical potentials. 
In this work, we keep the standard \cmf{} scalar self-interactions fixed in form, as described in~\cite{Papazoglou:1997uw,Kumar:2024owe,Cruz-Camacho:2024odu}, and focus on the baryon--meson coupling and on the quartic vector self-interactions. The vector self-interactions are particularly important for the density dependence of the EoS at and above $\nsat$, and for neutron star properties~\cite{Dexheimer:2015qha,Dexheimer:2018dhb}.

\subsection{Basics}

To describe the u, d, s quarks, and build hadrons from them, we write flavor matrices in the fundamental representation of $\suthree{}$. In this work we focus on hadronic matter, and we don't discuss quarks further. 
The singlet generator is
\begin{equation}
    T_0 = \frac{\mathbf{1}_{3\times 3}}{\sqrt{3}},
\end{equation}
and the octet generators are
\begin{equation}
    T_a = \frac{\lambda_a}{\sqrt{2}}, \qquad a=1,\ldots,8,
\end{equation}
where $\lambda_a$ are the Gell-Mann matrices.
The baryon octet is represented by a traceless matrix $B$, while the scalar and vector meson nonets are represented by matrices $X$ and $V_\mu$, respectively; in this work, Greek letters represent space-time indices, and we work with the metric $(+,-,-,-)$. These fields are expanded in the singlet plus octet basis spanned by $T_0$ and $T_a$. 
Under $\suthree{V}$, these fields transform as
\begin{equation}
    B \rightarrow h B h^\dagger, \qquad
    X \rightarrow h X h^\dagger, \qquad
    V_\mu \rightarrow h V_\mu h^\dagger,
\end{equation}
where matrix contractions onto flavor-adjoint indices are carried out with a Euclidean metric, and where $h\in SU(3)_V$~\cite{Coleman:1969sm,Callan:1969sn}. Therefore, meson interactions constructed from traces of products of $X$ and $V_\mu$ are invariant under the flavor transformation.

We work in the homogeneous, parity-even mean-field approximation. In this limit, meson fields are replaced by spacetime-independent expectation values, and the only nonzero vector mean fields are the temporal components.
The expectation values of the pseudoscalar Goldstone and axial-vector fields vanish in this approximation, although those fields remain part of the full nonlinear realization of chiral symmetry and enter the construction of the explicit symmetry-breaking terms~\cite{Papazoglou:1997uw}.
We also neglect flavor off-diagonal condensates. Therefore, only the diagonal singlet generator $T_0$ and the diagonal Cartan generators $ T_3$ and $T_8$ contribute to the meson fields. The relevant scalar and vector mean-field matrices can then be written as
\begin{equation}
    X = \mathrm{diag}\left( \frac{\sigma+\delta}{\sqrt{2}}, \frac{\sigma-\delta}{\sqrt{2}}, \zeta \right),
\end{equation}
and
\begin{equation}
    V = \mathrm{diag}\left( \frac{\omega+\rho}{\sqrt{2}}, \frac{\omega-\rho}{\sqrt{2}}, \phi \right),
    \label{eq:V_meson}
\end{equation}
where $V$ denotes the temporal component of $V_\mu=(V, \boldsymbol{0})$. The fields $\sigma$ and $\omega$ are the nonstrange isoscalar scalar and vector fields, $\delta$ and $\rho$ are the corresponding isovector fields, and $\zeta$ and $\phi$ are the corresponding fields that carry hidden strangeness. These physical fields correspond to linear combinations of the singlet and octet components of the scalar and vector nonets.

The full Lagrangian used within the chiral mean field model is laid out in detail in \cite{Cruz-Camacho:2024odu}.  Here we will focus specifically on the interaction terms, which are relevant for this work.  

\subsection{Baryon-meson and scalar interactions}
The interactions between baryons and mesons are built from the singlet and octet fields,  and can be expressed in terms of the nonet matrices as
\begin{align}
&\mathcal{L}_{\rm int}=-\sqrt{2}g_8^M \bigg[\alpha_M \mathrm{Tr}\left({\bar{B}O}[M,B]\right)\nonumber \\
&+(1-\alpha_M )\bigg( \mathrm{Tr}\left({\bar{B}O}\{M,B\}\right) -\frac{2}{3}\mathrm{Tr}\left({\bar{B}OB}\right)\mathrm{Tr}\left(M\right) \bigg) \bigg) \nonumber\\
&-\frac{1}{\sqrt{3}}g_1^M \mathrm{Tr}({\bar{B}OB})\mathrm{Tr}(M )\,,
\label{eq:L_BM}
\end{align}
where $O=1$ and $M=X$ for scalar mesons, and $O=\gamma^\mu$ and $M=V_\mu$  for vector mesons. The mixing of the singlet and octet is reflected in the baryon-meson couplings~\cite{Cruz-Camacho:2024odu,Papazoglou:1998vr,Miyatsu:2013yta,Colucci:2013pya}.
After evaluating \Cref{eq:L_BM} in the mean-field approximation, the baryon effective chemical potentials and masses take the form
\begin{equation}
\mu_i^*=\mu_i-g_{\omega i}\omega-g_{\rho i}\rho-g_{\phi i}\phi\,,
\label{eq:mu_eff}
\end{equation}
and
\begin{equation}\label{eq:m_eff}
m_i^\ast = g_{i\sigma}\sigma+g_{i\zeta}\zeta+g_{i\delta}\delta + m_0,
\end{equation}
where the baryon-meson couplings are related to the original scalar $g_8^S, g_1^S, \alpha_S$ and vector couplings $g_8^V, g_1^V, \alpha_V$, as discussed in the ~\Cref{sec:cmf-couplings}.  A bare mass term $m_0$ is added here to parametrize effects unaccounted for in the model that can contribute to  baryon mass generation. We consider a baryon bare mass term of the form
\begin{equation}
    \mathcal{L}_0 = -m_0 \Tr{\bar B B}.
\end{equation}
This term is a singlet contribution, as it contributes equally to all baryons.

The scalar meson interactions are constructed from the chiral invariant interactions of the form $I_n\equiv\mathrm{Tr}(X^n)$ for $n(>0) \in \Z$ and $I_0\equiv\mathrm{det}( X )$. A dilaton field $\chi$ is introduced to mimic the breaking of scale invariance~\cite{Schechter:1980ak,Carter:1995zi,Heide:1993yz,Sasaki:2011sd}. The scalar self-interaction Lagrangian is
\begin{equation}
\begin{split}
\label{eq:L_sc}
\mathcal{L}_{S}=&-\frac{1}{2}k_0\chi^2I_2+k_1 I_2^2+k_2 I_4+2k_3\chi I_0+k_{3N}\chi I_3 \\ 
 &+ \frac{\chi^4}{4}\ln \bigg( \frac{\chi^4}{\chi_0^4}\bigg)   + \frac{\epsilon}{3}\chi^4\ln\bigg(\frac{I_0}{\mathrm{det}\langle X _0\rangle}\bigg) - k_4\chi^4\,.
\end{split}
\end{equation}
We work in the frozen-glue-ball limit~\cite{Schechter:1980ak,Bonanno:2008tt}, where the field is frozen at its vacuum value $\chi=\chi_0$. We use the scalar self-interaction couplings as described in~\cite{Cruz-Camacho:2024odu}, and leave the exploration of the scalar sector to future work.

Explicit chiral symmetry breaking is also included in the model. Since the pseudoscalar Goldstone fields do not acquire expectation values in the mean-field approximation, the explicit symmetry-breaking term reduces to
\begin{equation}\label{eq:LBS_mf}
\mathcal{L}_{\rm SB}=-\Tr{A_{SB} X},\,
\end{equation}
where the constant flavor matrix $A_{SB}$ is given by $A_{SB}=\frac{1}{\sqrt{2}} \mathrm{diag}(m_\pi^2f_\pi, \, m_\pi^2f_\pi, \, 2m_K^2f_K-m_\pi^2f_\pi)$~\cite{Koch:1997ei}, with $f_\pi$ and $f_K$ being the decay constants of pions and kaons. In the full non-linear realization, this term ensures that the partially conserved axial current, or PCAC, relations, are satisfied, and gives the Goldstone bosons their explicit symmetry-breaking masses. In the mean-field approximation, although the Goldstone bosons do not appear as independent bulk mean fields, their contribution through the explicit symmetry-breaking term still remains in \Cref{eq:LBS_mf}.

\subsection{The quartic vector self-interaction}
\label{sec:vector-couplings}

The density dependence of the \cmfpp EoS is strongly affected by the vector sector. In the mean field approximation, the quadratic vector contribution is
\begin{equation}
    \mathcal{L}_{V^2} =\frac{1}{2}\left(m_\omega^2\omega^2+m_\phi^2\phi^2+m_\rho^2\rho^2\right),
\end{equation}
where we fix the meson masses according to the values specified in Table IX of~\cite{Cruz-Camacho:2024odu}.

At fourth order, the flavor structure allows for several trace invariants to be constructed from the vector nonet matrix $V$. Generically, all possible vector self-interactions, constrained by the flavor structure of the model, are described by
\begin{equation}
\begin{split}
    \label{eq:couplings-definitions}
    \mathcal{L}_{V^4} &= g_4^{04} \Tr{V^4} + g_4^{13} \Tr{V} \Tr{V^3} \\
    & + g_4^{22} \left[\Tr{V}\right]^2 \Tr{V^2} 
    + g_4^{220} \left[\Tr{V^2}\right]^2,
\end{split}
\end{equation}
When expressed in terms of the physical fields $(\omega,\rho,\phi)$, the trace structure of the interactions generates multiple interaction channels, such as $\omega^4$, $\omega^2\rho^2$, $\phi^4$, and also other terms, which are not independent but constrained by the underlying $\suthree{V}$ symmetry. 
A fifth quartic interaction term, $g_4^{40}[\Tr{V}]^4$, can also be included. However, for a $3\times3$ matrix $V$, the Cayley--Hamilton theorem relates the five quartic trace structures to each other, leaving only four linearly independent invariants~\cite{Cata:2007ns}. We therefore choose the basis in \Cref{eq:couplings-definitions}, omitting the \([\Tr{V}]^4\) term.

To gain physical insight into the proposed vector self-interaction of \Cref{eq:couplings-definitions}, we can write the vector nonet matrix in the diagonal singlet--octet basis
\begin{equation}
    V = v_0 T_0 + v_3 T_3 + v_8 T_8,
\end{equation}
such that the components $V=\mathrm{diag}(V_u, V_d, V_s)$ are
\begin{equation}
    \begin{split}
    V_u &= \frac{v_0}{\sqrt{3}} + \frac{v_8}{\sqrt{6}} + \frac{v_3}{\sqrt{2}}, \\
    V_d &= \frac{v_0}{\sqrt{3}} + \frac{v_8}{\sqrt{6}} - \frac{v_3}{\sqrt{2}}, \\
    V_s &= \frac{v_0}{\sqrt{3}} - \frac{2\,v_8}{\sqrt{6}}
\end{split}
\end{equation} 
Under ideal mixing, these are expressed in terms of the physical mesons ($\omega, \phi, \rho$) as (see \Cref{eq:V_meson})
\begin{equation}
\begin{split}
    v_0 &= \frac{\sqrt{2}\,\omega + \phi}{\sqrt{3}}, \\
    v_3 &= \rho, \\
    v_8 &= \frac{\omega - \sqrt{2}\phi}{\sqrt{3}}.
\end{split}
\end{equation}

Substituting these expressions into \Cref{eq:couplings-definitions}, the quartic interaction can be written as
\begin{equation}
\begin{split}
    \mathcal{L}_{V^4} &= \mathcal{C}_{0}\, v_0^4 + \mathcal{C}_{8}\left(v_3^2 + v_8^2 \right)^2+ \mathcal{C}_{08}\,v_0^2\left(v_3^2 + v_8^2\right)\\ 
    &+ \mathcal{C}_{038}\,\left( v_0\, v_3^2\, v_8 - \frac{1}{3} v_0 v_8^3\right),
\end{split}
\end{equation}
The four coefficients have a transparent interpretation: $\mathcal{C}_0$ controls the pure singlet quartic term, $\mathcal{C}_8$ controls the pure octet quartic term, $\mathcal{C}_{08}$ controls the quadratic singlet--octet mixing, and $\mathcal{C}_{038}$ controls the singlet times cubic-octet invariant. They are related to the trace-basis couplings by
\begin{equation}
    \label{eq:physical-couplings-defs}
    \begin{split}
        &\mathcal{C}_0 = g_4^{04}/3  + g_4^{13} + 3 g_4^{22} + g_4^{220} \\ 
        &\mathcal{C}_8 =g_4^{04}/2 + g_4^{220} \\
        &\mathcal{C}_{08} = 2 g_4^{04} + 3 g_4^{13} + 3g_4^{22} + 2g_4^{220} \\
        &\mathcal{C}_{038} =  \left( 4 g_4^{04} + 3 g_4^{13}\right)/\sqrt{2} \\
    \end{split}
\end{equation}
If the omitted term $g_4^{40}[\Tr{V}]^4$ is included, it only shifts the pure singlet coefficient,
\begin{equation}
\begin{split}
\bar g_4^{04} &= g_4^{04}+6g_4^{40},\\
\bar g_4^{13} &= g_4^{13}-8g_4^{40},\\
\bar g_4^{22} &= g_4^{22}+6g_4^{40},\\
\bar g_4^{220} &= g_4^{220}-3g_4^{40}.
\end{split}
\end{equation}

As a final observation, we note that writing down the explicit interactions between the $\omega, \phi$ and $\rho$ mesons, we have 9 non-vanishing  terms $\{\omega^4, \phi^4, \rho^4, \omega^2 \rho^2, \phi^2 \rho^2, \omega^3\phi, \omega \phi^3, \omega^2 \phi^2, \omega \phi \rho^2\}$, whose corresponding couplings can be written as a linear combination of the four $g_4^{ij}$ couplings.  These expressions are given in~\Cref{app:phys_couplings}.
Therefore, the \cmfpp flavor structure does not allow one to tune each physical channel independently. Instead, the high-density behavior of the EoS is governed by correlated combinations of the couplings $g_4^{04}$, $g_4^{13}$, $g_4^{22}$, and $g_4^{220}$. 
This structure is the main motivation for the inference of this paper. Instead of assuming priors on the independent physical terms, we assign independent priors to each $g_4^{ij}$ and infer which combinations are compatible with vacuum and nuclear-matter constraints.

The densities, energy density, and pressure of the baryons are given by the typical Fermi-Dirac expressions with mean-field interactions that can be computed from the energy-momentum tensor. The field equations are obtained by minimizing the thermodynamic potential at fixed temperature and chemical potentials (see~\cite{Cruz-Camacho:2024odu} for the expressions).

\section{Chiral Mean Field Parameters}
\label{sec:cmf-couplings}

The \cmf{} parameters explored in this work can be naturally split into vacuum and in-medium sectors, allowing the parameter space to be split as
\begin{equation}
    \theta_{\rm \cmf} = \theta_{\rm vac} \cup \theta_{\rm med},
\end{equation}
where $\theta_{\rm vac}$ is a list of vacuum parameters and $\theta_{\rm med}$ a list of in-medium parameters. Each of these leads to observables that can be classified into two sets, one that depends only on the vacuum parameters $F_{\rm vac} = F_{\rm vac}(\theta_{\rm vac})$, and another that depends on both vacuum and in-medium parameters $F_{\rm sat} = F_{\rm sat}\left(\theta_{\rm vac}, \theta_{\rm med} \right)$.  This separation is useful because the vacuum masses depend only on the scalar baryon--meson couplings, whereas saturation properties depend on both the scalar sector and the vector-sector parameters, and will simplify the Bayesian inference in \Cref{sec:bayesian-theory}.
In the subsections below, we describe the parameters and the associated observables related to each sector.

\subsection{Vacuum sector}
\label{sec:cmf_params_vacuum}
In a chiral model, the baryon masses are generated dynamically by the scalar mesons, which take non-vanishing expectation values in vacuum, $\sigma = \sigma_0$, $\zeta = \zeta_0$ and $\delta = 0$, arising from solving the equations of motion with vanishing source terms. The expression for the baryon masses,~\Cref{eq:m_eff}, in vacuum becomes
\begin{equation}
    \label{eq:baryon-masses}
    m_{i}^{\rm vac} = m_{0} + g_{\sigma i} \sigma_0 + g_{\zeta i} \zeta_0,
\end{equation}
where the meson expectation values in vacuum are functions of the pion and kaon decay constants, and we take them to have known values according to~\cite{Papazoglou:1998vr,FlavourLatticeAveragingGroupFLAG:2024oxs}\footnote{Changing these constants requires re-solving the vacuum equations of motion, and doing so would require other choices of the scalar self interaction couplings; see Eq.~\eqref{eq:L_sc} and surrounding discussion.}
Therefore, the observables that can be computed in the vacuum sector are the octet baryon masses\footnote{We do not include the baryon decuplet in this work.}
\begin{equation}\label{eq:F_vac}
    F(\theta_{\rm vac}) = \left( m_N, m_\Lambda, m_\Sigma, m_\Xi\right).
\end{equation}
Here we use the four isospin-averaged baryon masses, namely, the nucleon mass $m_{N}$, the $\Lambda$ mass $m_{\Lambda}$, the $\Sigma$ mass $m_{\Sigma}$ and the $\Xi$ mass, $m_{\Xi}$. 
While experiments show \cite{ParticleDataGroup:2026aaa} that physical masses of baryons in the same isospin multiplet are split, these effects arise from electromagnetic interactions and from explicit isospin symmetry breaking $m_u \neq m_d$ \cite{BMW:2014pzb}, which are not included in the \cmfpp vacuum.

The vacuum parameters $\theta_{\rm vac}$ are the subset of parameters on which the masses depend. These are not the 16 scalar couplings to the baryons, $g_{i\sigma}$ and $g_{\zeta i}$, as those are functions of the original couplings that enter the Lagrangian, \Cref{eq:L_BM}.  Instead, we infer a set of parameters which determine the scalar-baryon couplings:   
\begin{equation}\label{eq:theta_vac}
    \theta_{\rm vac} = \left( g_1^S, g_8^S, \alpha_S\right),
\end{equation}
defined in ~\Cref{eq:L_BM}.
Explicitly, the scalar meson couplings to the nucleon are 
\begin{align}
    \label{eq:nucleon-sigma-coupling}
    g_{\sigma N} &= \frac{1}{3}g_8^S\quant{4\alpha_S-1} + \sqrt{\frac{2}{3}}g_1^S,\\
    \label{eq:nucleon-zeta-coupling}
    g_{\zeta N} &= -\frac{\sqrt{2}}{3} g_8^S\quant{4\alpha_S-1} + \frac{1}{\sqrt{3}} g_1^S,
\end{align}
for the $\Lambda$ they are
\begin{align}
g_{\sigma\Lambda} &=
\frac{2}{3} g_8^S(\alpha_S-1)
+ \sqrt{\frac{2}{3}}\,g_1^S, \\
g_{\zeta\Lambda} &=
-\frac{2\sqrt{2}}{3} g_8^S(\alpha_S-1)
+ \frac{1}{\sqrt{3}}g_1^S,
\end{align}
for the $\Sigma$ they are
\begin{align}
g_{\sigma\Sigma} &=
-\frac{2}{3} g_8^S(\alpha_S-1)
+ \sqrt{\frac{2}{3}}\,g_1^S, \\
g_{\zeta\Sigma} &=
\frac{2\sqrt{2}}{3} g_8^S(\alpha_S-1)
+ \frac{1}{\sqrt{3}}g_1^S,
\end{align}
and for the $\Xi$ they are
\begin{align}
g_{\sigma\Xi} &=
-\frac{1}{3} g_8^S(2\alpha_S+1)
+ \sqrt{\frac{2}{3}}\,g_1^S,\\
g_{\zeta\Xi} &=
\frac{\sqrt{2}}{3} g_8^S(2\alpha_S+1)
+ \frac{1}{\sqrt{3}}g_1^S.\label{eq:xi-zeta-coupling}
\end{align}

The masses also depend on the bare mass $m_0$, although, in practice, $m_0$ and $g_1^S$ are degenerate in the vacuum: both $g_1^S$ and $m_0$ are singlet contributions to the masses, and therefore the vacuum masses contribute identically to all baryons. To be precise, the vacuum masses depend on $m_0$ and $g_1^S$ only through the combination
\begin{equation}    
m_0 + g_1^S\left(\sqrt{2/3}\sigma_0 + \sqrt{1/3} \zeta_0\right).
\end{equation}
This implies that, given four target values for the vacuum masses $F_{\rm vac}$, we cannot independently determine $m_0$ and $g_1^S$. We will take $m_0=0$ in the main analysis of this paper and, in \Cref{app:m0-degenerate-sampling}, we will analyze the expanded set of vacuum parameters
\begin{equation}
    \theta_{\rm vac}^{(m_0)} = \left( g_1^S, g_8^S, \alpha_S, m_0\right).
\end{equation}

\subsection{In-medium sector}
\label{sec:cmf_params_medium}

We now discuss the observables and parameters of the \cmf{} model that are relevant for dense matter. In this work, we use \emph{in-medium} to refer to finite-density calculations in infinite nuclear matter. Near saturation, the energy per baryon, measured relative to the nucleon mass, is conventionally organized as a Taylor expansion in baryon number density, $n_B$, and isospin asymmetry, $1-2Y_Q$, where $Y_Q$ is the proton fraction\footnote{This approach only works when strangeness is assumed to be zero, which is reasonable at $\nsat$ in the absence of hypernuclei, otherwise the isospin asymmetry must take strangeness into account, see \cite{Yang:2025wop,Danhoni:2025qpn}}. Keeping the standard quadratic dependence on isospin asymmetry, and expanding around symmetric nuclear matter at saturation density, $\nsat$, we write
\begin{align}
    \label{eq:energy-expansion}
    E(n_B,Y_Q) = E_{\SNM}(n_B) + (1-2Y_Q)^2 S(n_B),
\end{align}
with
\begin{equation}
\label{eq:snm-expansion}
    E_{\SNM}(n_B) = \Esat + \half K_{0} \quant{\frac{n_B-\nsat}{3\nsat}}^2,
\end{equation}
and
\begin{equation}
\label{eq:symmetry-energy-expansion}
    S(n_B) 
    \equiv  \Esym + \Lsym \quant{\frac{n_B-\nsat}{3\nsat}}
\end{equation}
where higher order terms have no existing experimental constraints such that we do not consider them here. 

Here $\Esat$ is the binding energy of symmetric nuclear matter at saturation density relative to the nucleon mass, $K_0$ is the incompressibility, $\Esym$ is the symmetry energy at saturation, $\Lsym$ is its slope. When only up to quadratic terms are kept in isospin asymmetry, $\Esym = E_{\PNM}-E_{\SNM}$ at saturation, with $E_{\PNM}=E(\nsat,Y_Q=0)$. We work at zero temperature and neglect strangeness, since strange degrees of freedom are not expected to affect nuclear matter near saturation.

The quantities in \Cref{eq:snm-expansion,eq:symmetry-energy-expansion} are often called the ``saturation parameters,'' or ``saturation properties,'' of nuclear matter. Here, we treat them as the \emph{saturation observables}, or \emph{in-medium observables}, used to calibrate the in-medium behavior of the \cmf{} EoS:
\begin{equation}\label{eq:F_sat}
    F_{\rm sat} = \left(\nsat, \Esat, K_{0}, \Esym, \Lsym \right).
\end{equation}
These observables are derived from the \cmf{} model, but they are not available as simple closed-form functions of the parameters. For each parameter choice, the equations of motion must be solved in medium, and the resulting $E_{\SNM}(n_B)$ and $E_{\PNM}(n_B)$ curves must be fit to the forms in \Cref{eq:snm-expansion,eq:symmetry-energy-expansion}. This is unlike the vacuum masses discussed in \Cref{sec:cmf_params_vacuum}, which can be computed analytically. Computing $F_{\rm sat}$ is therefore the expensive part of the model evaluation.

The in-medium observables are functions not only of the vacuum parameters, \Cref{eq:theta_vac}, but also of the vector meson coupling to the baryons and the vector meson self-couplings, i.e. $F_{\rm sat} = F_{\rm sat}(\theta_{\rm vac}, \theta_{\rm med})$, with 
\begin{equation}\label{eq:theta_med}
    \theta_{\rm med} = \left(g_{N\omega}, g_{N\rho}, g_4^{04}, g_4^{13}, g_4^{22}, g_4^{220}\right).
\end{equation}
Now, we discuss our strategy for determining the model parameters, $\theta_{\rm CMF}$.

\section{Bayesian Hierarchical Inference}
\label{sec:bayesian_analysis}

The \cmfpp construction described above defines two model maps. The vacuum map, given by \Cref{eq:F_vac},  relates the scalar-sector parameters to the isospin-averaged baryon masses, and the in-medium map, given by \Cref{eq:F_sat}, relates the entire parameter space to the saturation properties of cold nuclear matter. 
Hierarchical Bayesian inference naturally provides a framework for incorporating independent constraints on a model while faithfully including uncertainty in measurements.  
In this section, we define the effective target values used for these observables, the likelihood, and the hierarchical sampling strategy, with the aim to infer regions of \cmf{} parameter space that are consistent with measurements of nuclear matter.

\subsection{Hierarchical Bayesian theory}
\label{sec:bayesian-theory}

To determine the \cmf{} parameters $\theta_{\rm CMF}$, we compare two sets of model predictions to effective target values. The first set is the vacuum map, $F_{\rm vac}(\theta_{\rm vac})$, which gives the isospin-averaged baryon masses. The second is the saturation map, $F_{\rm sat}(\theta_{\rm vac},\theta_{\rm med})$, which gives the in-medium observables near nuclear saturation. We denote the corresponding target values by $d_{\rm vac}$ and $d_{\rm sat}$.

The uncertainties assigned to these targets should be interpreted as effective tolerances, not as purely experimental errors. They include both measurement uncertainty and the systematic error associated with using the \cmf{} model to describe averaged baryon masses and empirical saturation properties. For example, the \cmf{} model predicts an isospin-averaged nucleon mass, but not the proton-neutron mass splitting in vacuum. We therefore, compare the model to an effective nucleon mass with an uncertainty that reflects this modeling limitation.

Given these targets and uncertainties, the inference follows from Bayes' theorem,
\begin{equation}
    \label{eq:bayes-complete}
    P(\theta_{\rm CMF}|d_{\rm vac}, d_{\rm sat}) =
    \frac{\mathcal{L}(d_{\rm vac}, d_{\rm sat} | \theta_{\rm CMF})
    \,\pi(\theta_{\rm CMF})}{\mathcal Z(d_{\rm vac}, d_{\rm sat})},
\end{equation}
where $\pi(\theta_{\rm CMF})$ is the prior and $\mathcal L(d_{\rm vac}, d_{\rm sat}|\theta_{\rm CMF})$ is the likelihood.  The posterior, $P(\theta_{\rm CMF}| d_{\rm vac}, d_{\sat})$, is defined by \Cref{eq:bayes-complete}. The evidence, $\mathcal Z(d_{\rm vac}, d_{\rm sat})$, is defined so that the posterior distribution is normalized.  We take the likelihood to be Gaussian in the differences between the target values and the corresponding \cmfpp predictions. The priors are specified in \Cref{sec:priors}.

\begin{table*}[thb]
    \centering
    \caption{
    Target baryon mass values and uncertainties used in the first stage of the hierarchical analysis (left) and Gaussian mean and uncertainty for the saturation observables used in the second stage (right). All quantities are in MeV, except $\nsat$, which is in $\mathrm{fm}^{-3}$.}
    \renewcommand{\arraystretch}{1.5}
    \begin{tabular}{|c|c|c|c|}
    \hline
    Particle & Mean [$\mev{}$] & Uncertainty [$\mev{}$] & Represents\\
    \hline
       N  & $937.2$ & $1$ & n, p  \\
       \hline
         $\Lambda$ &$1115$& $7$ &$\Lambda$\\
         \hline
         $\Sigma$ & $1202$& $7$ &$\Sigma^-, \Sigma^0, \Sigma^+$\\
         \hline
         $\Xi$ &$1336$ & $7$ & $\Xi^0, \Xi^{-}$ \\
         \hline
    \end{tabular}
    \qquad \qquad
    \begin{tabular}{|c|c|c|}
        \hline
        Property & Mean & Uncertainty  \\
        \hline
         $\nsat\, [\mathrm{fm}^{-3}]$ & 0.155  & 0.005 \\
         \hline
         $\Esat\, [\mev]$  & -16.0 & 0.6\\
         \hline
          $K_0\,[\mev]$ & 240 & 100\\
          \hline
         $\Esym\,[\mev]$ &  32.0 & 2.0 \\
         \hline
        $\Lsym\,[\mev]$ & 60.0 & 20.0\\
         \hline
    \end{tabular}
    \label{tab:mass-measurements}
\end{table*}

The hierarchical structure enters because the two maps depend on different subsets of parameters. The vacuum masses depend only on $\theta_{\rm vac}$, while the saturation observables depend on both $\theta_{\rm vac}$ and $\theta_{\rm med}$. Thus, the vacuum data can be used first to constrain $\theta_{\rm vac}$, and the resulting posterior can then be used as input to the in-medium inference.

This is a special case of a more general simplification. Suppose an observable $d_i$ has a model prediction $F_i$ that depends only on a subset of the full parameter vector, $\theta_{(i)}\subseteq\theta$. Then the corresponding likelihood factor depends only on that subset,
\begin{equation}
    \label{eq:individual-likelihood-reduced-dependence}
    \mathcal L(d_i | \theta) = \mathcal L \big(d_i | \theta_{(i)}\big).
\end{equation}
If the uncertainties on different observables are treated as independent, the full likelihood factorizes into a product of such terms. This reduces the effective dimensionality of the sampling problem. In practice, this matters: even a five-dimensional grid with $\sim100$ samples per dimension would require $\mathcal O(10^{10})$ likelihood evaluations. Stochastic samplers avoid a full grid, but they still become more expensive as the number of dimensions grows.

For the vacuum sector, assuming independent Gaussian uncertainties, the likelihood is
\begin{equation}
\label{eq:L_vac}
\log \mathcal L_{\rm vac} =
-\frac{1}{2} \sum_{i\in{\rm vac}} \left[
\frac{F_{{\rm vac},i}(\theta_{\rm vac})-d^{\rm vac}_i}
{\sigma^{\rm vac}_i}
\right]^2 ,
\end{equation}
where $i\in{\rm vac}$ runs over the vacuum masses of $\{N,\Lambda,\Sigma,\Xi\}$. The target values are listed in \Cref{tab:mass-measurements}. The saturation likelihood is
\begin{equation}
\label{eq:L_sat}
\log \mathcal L_{\rm sat} =
-\frac{1}{2} \sum_{i\in{\rm sat}} \left[
\frac{F_{{\rm sat},i}(\theta_{\rm CMF})-d^{\rm sat}_i}
{\sigma^{\rm sat}_i}
\right]^2 ,
\end{equation}
where $i\in{\rm sat}$ runs over the saturation observables used as constraints. In this work, we constrain $\nsat$, $\Esat$, $K_0$, $\Esym$, and $\Lsym$.
The corresponding target values are listed in \Cref{tab:mass-measurements}.

This independent-Gaussian likelihood is an approximation for exploratory inference. It neglects correlations among empirical saturation quantities, especially between $\Esym$ and $\Lsym$~\cite{Newton:2020jwn,Holt:2018uug,Margueron:2018eob}. The posterior should therefore be interpreted as a map of plausible parameter regions under these effective constraints, not as a precision statistical determination of the model parameters.

\subsection{Target values and priors}
\label{sec:priors}

The numerical target values entering the likelihoods of \Cref{eq:L_vac,eq:L_sat} are listed in \Cref{tab:mass-measurements}. As discussed above, the quoted uncertainties should be interpreted as effective $1$--$\sigma$ tolerances, not as purely experimental errors.

For the vacuum sector, we use fiducial isospin-averaged baryon masses chosen for consistency with previous \cmf{}parametrizations. These values should be understood as effective calibration targets, rather than as the most up-to-date PDG central values~\cite{ParticleDataGroup:2024cfk}. We assign an uncertainty of $1\,\mev{}$ to the nucleon mass and $7\,\mev{}$ to each hyperon mass. These uncertainties are dominated by model systematics, including the fact that the vacuum of the model does not resolve the mass splittings inside an isospin multiplet.

For the saturation observables, we synthesize representative values and uncertainties from theoretical and experimental constraints on dense matter, as summarized in \Cref{tab:mass-measurements}. These choices are broadly consistent with a range of studies~\cite{Lattimer:2012xj, Reed:2021nqk, Adhikari:2021phr, CREX:2022kgg, ParticleDataGroup:2024cfk} and references therein. We do not attempt to construct a complete covariance model for these empirical constraints, because our goal is exploratory: to identify plausible regions of the \cmf{} parameter space. The quoted uncertainties are therefore deliberately effective and, in some cases, conservative. In particular, although $\Esym$ and $\Lsym$ are  correlated within many models and constraints~\cite{Kortelainen:2010hv,Lattimer:2012xj, Nazarewicz:2013gda, Holt:2018uug, Newton:2020jwn, Reed:2021nqk}, we treat the saturation observables as independent in the likelihood. This is important because we explore the vector sector interactions in this work, which strongly affect this correlation~\cite{Dexheimer:2018dhb}. We leave the exploration of correlations in the \cmf{} model to future work.

We must also specify priors for the model parameters. For the vacuum parameters, the baryon-mass targets and the relations between \Cref{eq:nucleon-zeta-coupling,eq:xi-zeta-coupling} are sufficient to choose finite prior ranges that contain the relevant likelihood support. This is not automatic, because the \cmf{} model is phenomenological. Even if one could place constraints on particular low-energy interactions using, for example, chiral perturbation theory, the \cmf{} couplings absorb the effects of many interactions that are not included explicitly, such as multiple-pion exchange. In the main analysis, we fix $m_0=0$, consistent with the C1, C2, and C3 schemes of~\cite{Dexheimer:2015qha}. The corresponding prior ranges for $\theta_{\rm vac}$ are shown in \Cref{tab:priors-no-m-zero}.
\begin{table}[]
    \renewcommand{\arraystretch}{1.5}
    \centering
    \begin{tabular}{|c|c|c|}
    \hline
         Parameter & Minimum & Maximum \\
         \hline
         $g_1^S$&-14.3 & -0.3\\
         \hline
         $g_8^S$ & -5.3 & 0.7\\
         \hline
         $\alpha_S$ & 0 & 2.0\\
         \hline
    \end{tabular}
    \caption{Priors on the vacuum-sector parameters for the fixed-$m_0=0$ analysis.}
    \label{tab:priors-no-m-zero}
\end{table} 
We also consider an extended vacuum sector in which $m_0$ is allowed to vary. This case is discussed in \Cref{app:m0-degenerate-results}.

For the in-medium parameters $\theta_{\rm med}$, the situation is different. There are fewer constrained observables than parameters, and the high-likelihood region cannot be enclosed efficiently by simple parameter-by-parameter prior bounds. The vector-sector parameters, especially the quartic self-couplings, control correlated combinations of physical interaction channels, so the region compatible with saturation properties is not known before sampling. We therefore determine the in-medium prior region iteratively, as described in \Cref{sec:iterative-emulation}.

This also makes direct sampling expensive. The saturation likelihood is narrow compared to the range of model predictions allowed by broad priors, and faithful stochastic sampling can require millions of likelihood evaluations. Each likelihood evaluation requires an in-medium model evaluation, including the solution of the coupled field equations on a grid. Even if one such evaluation took only one second, a direct inference would still be impractical. In the next section, we therefore introduce neural-network emulators for the map from \cmf{} parameters to saturation observables, together with an iterative strategy for expanding the training data in the relevant region of parameter space.

\section{Emulator of saturation properties and Iterative Inference Scheme}
\label{sec:emulator-and-iterative-scheme}

We implement a fairly simple feedforward neural network to approximate the mapping

\begin{equation}
    \label{eq:function-sat-props-schematic}
    F_{\rm sat}: \{\theta_j\} \mapsto (\nsat, \Esat, K_0, \Esym, \Lsym).
\end{equation}
As we will demonstrate, we find that, with only 2 hidden layers of $\mathcal O(100)$ neurons, we are able to achieve reasonable fits to the model given sufficient training data in the region of interest.  We discuss the neural network emulator in \Cref{sec:neural-network-details}.  We adopt a strategy of simultaneously varying the priors in our inference, and expanding the range of training data for the emulator based on noisy samples from the previous iteration. Through this procedure we sequentially obtain better fits; see \Cref{sec:iterative-emulation}. We stress that emulation is only necessary for calculations at finite density, vacuum masses are computed exactly within the \cmf{} framework outlined in Sec.~\ref{sec:cmf-descripton}.

\subsection{Details of the Neural Network Emulator}
\label{sec:neural-network-details}

We use a feed-forward neural network implemented in {\sc pytorch}. In the main analysis, the network has nine nontrivial inputs, corresponding to the parameters $\theta_{\rm vac} \cup \theta_{\rm med}$ described in \Cref{sec:cmf_params_vacuum,sec:cmf_params_medium}.  The outputs are the five saturation observables used in the likelihood, $(\nsat,\Esat,K_0,\Esym,\Lsym)$.  The emulator has two dense hidden layers, each with 128 neurons, and is trained with the native {\sc pytorch} \texttt{Adam} optimizer and mean-squared-error loss, \texttt{MSELoss}.  Full details of the configuration are provided in the accompanying data release.

Training data are generated by direct \cmfpp calculations, using the software stack summarized in \Cref{sec:computational-considerations}.    In each iteration, we use $85\%$ of the generated samples for training and $15\%$ for validation.

A practical complication is that broad \cmf{} priors include many parameter choices whose saturation observables are extremely far from the empirical region.  If these values are used directly, they can dominate the mean-squared-error loss, even though the precise magnitude of such an excursion is irrelevant for the inference.  For example, once a model predicts a value of $\nsat$ or $\Lsym$ far outside the region where the likelihood is appreciable, it matters much more that the emulator identifies the point as ``too large'' or ``too small'' than that it reproduce exactly how far outside the region the point lies.

A natural solution to this is to specify ahead of time a  ``physical range"  for each parameter, with maximum and minimum values reflecting rough
``limits" on which the model reproduces qualitatively the correct physics.
If a computed value lies above or below the adopted range, we replace it by the corresponding upper or lower bound for the purposes of training.
We term this process ``clamping", and we display training targets for each saturation observable to the ranges shown in \Cref{tab:saturation-clipping}.    This should be understood as a preprocessing step on the emulator targets, not as a modification of the \cmfpp calculation itself.  We do not simply discard these samples, because then the emulator would receive no information about large regions of parameter space that are clearly incompatible with saturation constraints.  Clamping retains the useful information, namely that the point lies safely outside the inference region and on which side of it, while preventing the optimizer from spending most of its capacity fitting the exact numerical value of unphysical outliers.

The price of this choice is that the emulator should not be interpreted as an accurate extrapolation outside the clamping ranges.  This is acceptable for our purposes because the boundaries in \Cref{tab:saturation-clipping} are chosen to lie well outside the uncertainty ranges used in the likelihood.  Within the region relevant for the posterior, the clamping does not alter the training labels.  We choose the ranges empirically, narrowing them until large outliers no longer dominate the training loss.  We discuss the resulting emulator performance in \Cref{sec:emulator-performance}.

\begin{table}[]
    \centering
    \renewcommand{\arraystretch}{1.5}
    \begin{tabular}{|c|c|c|}
        \hline
        property & lower bound & upper bound  \\
        \hline
         $\nsat\, [\mathrm{fm}^{-3}]$ & 0.0& 0.4\\
         \hline
         $\Esat\, [\mev]$  & -45& 15\\
         \hline
          $K_0\,[\mev]$ & -150 & 650\\
          \hline
         $\Esym\,[\mev]$ &  -10 & 50\\
         \hline
        $\Lsym\,[\mev]$ & -50& 150\\
         \hline
    \end{tabular}
    \caption{Clamping ranges used when training the emulator. Values outside these intervals are set to the corresponding lower or upper bound for the purposes of emulator training. All quantities are in MeV except $\nsat$, which is in $\mathrm{fm}^{-3}$.}
    \label{tab:saturation-clipping}
\end{table}

\subsection{Iterative emulation scheme}
\label{sec:iterative-emulation}
As described in \Cref{sec:bayesian-theory}, the \cmf{} parameters are not well-constrained \emph{a priori}.    It is known, however, that certain configurations of parameters produce essentially physical EoSs, which satisfy constraints on saturation properties and produce, e.g. $\sim2\,M_{\odot}$ stars\footnote{In this work, we do not use the properties of neutron stars in our likelihood, though we do verify that our posteriors are consistent with $2\,M_{\odot}$ stars if they are completely composed of nuclear matter.  See \Cref{sec:discussion}.}.  Here, we leverage the configuration of the C1 parametrization of \cmf~\cite{Dexheimer:2015qha}, under the hypothesis that points in parameter space that are sufficiently close to this parameter configuration may also produce physical EoSs.  We therefore initially sample from a prior that reflects points ``close" to C1, in the full parameter space that we use here, given by~\Cref{eq:couplings-definitions}.   We show the initial prior distribution we use on the vector couplings in \Cref{tab:vector-priors-initial}.
\begin{table}[]
    \centering
    \renewcommand{\arraystretch}{1.5}
    \begin{tabular}{|c|c|c|}
        \hline
        Parameter & mean & standard deviation \\
        \hline
        $g_4^{04}$   & 116.8 & 40.0 \\
        \hline
        $g_{N\omega}$ & 13.66 & 2.0 \\
        \hline
        $g_{N\rho}$   & 4.94 & 1.0 \\
        \hline
        $g_4^{13}$   & 0.0 & 10.0 \\
        \hline
        $g_4^{22}$   & 0.0 & 10.0 \\
        \hline
        $g_4^{220}$  & 0.0 & 10.0 \\
        \hline
    \end{tabular}
    \caption{Gaussian priors for \cmf{} parameters in the vector sector used in the initial prior set.}
    \label{tab:vector-priors-initial}
\end{table}

We additionally use a multivariate Gaussian fit to the posterior of the inference described in \Cref{sec:priors}, with priors explicitly shown in \Cref{tab:priors-no-m-zero}.    For initial training data, several avenues are possible.  First, one may simply sample the full joint prior and evaluate directly the model parameters using the \cmf{} model itself.  This procedure may be inefficient, since most of the prior volume is unphysical.  In practice, we use the output of prior sampling runs in our production runs to improve the number of physically useful samples the emulator has access to. 

The iterative algorithm then proceeds as follows: 
\begin{enumerate}
    \item An emulator for the saturation properties is trained on all of the available training data. 
    \item The sampler is run with the given priors using the emulator to evaluate saturation properties, which effectively determines the likelihood given the ``target values" of \Cref{tab:mass-measurements}.
    \item The posterior is used to (a) select new training points by adding random noise to posterior samples\footnote{We do not use posterior samples directly to generate new training data because we still need the emulator to be somewhat effective even in regions with low posterior probability.  Therefore, our procedure produces new training points that are in the ``neighborhood" of the posterior, but distributed more broadly than the posterior itself.} and evaluating their saturation properties using the full \cmf{} model and (b) adjust the range of the priors, which are set to have widths some multiple (which depends on the iteration) of the prior iteration posterior width.  
    \item New training data consists of all previous training data, plus the new \cmf{} parameter samples drawn and evaluated. We return to (1.) and retrain the emulator, and continue the process until a desired termination criteria is met.   
\end{enumerate}

The key detail of this algorithm is that new training data must be generated in the region that is to be sampled for the sampler to accurately produce a posterior estimate that matches what it would get if it used the underlying model. However, we also find that, at each iteration, the number of training samples needed to produce plausible fits is not exceptionally large, of order a few hundred if the prior range is expanded by order $1\sigma \approx 30\%$ the width of the posterior.  We attribute this to the fact that near the boundary of the prior, the posterior is suppressed by the prior (since all our priors are Gaussian).  Therefore, even if the likelihood is mismodeled, samples with few training points still do not contribute too much to the posterior at any iteration.  Those regions that are relevant \emph{a posteriori} will preferentially receive additional coverage from the next set of training samples, according to the algorithm. Therefore, the algorithm regulates mismodeling due to the use of the emulator. We find that, after several iterations, regions of parameter space that are persistently in the posterior are very well emulated.   We quantify all of this below, in \Cref{sec:results}.

Finally, for the runs presented in this work, we repeat the entire iterative scheme twice, using all of the training data generated during the first run as initial training data for the second run.  We find this improves the performance of the emulator, and anecdotal evidence suggests that repeating this procedure would increase the precision of the emulator. We discuss this as well in \Cref{sec:results}.

\subsection{Computational considerations}
\label{sec:computational-considerations}

\begin{figure}
    \centering
    \includegraphics[width=0.99\linewidth]{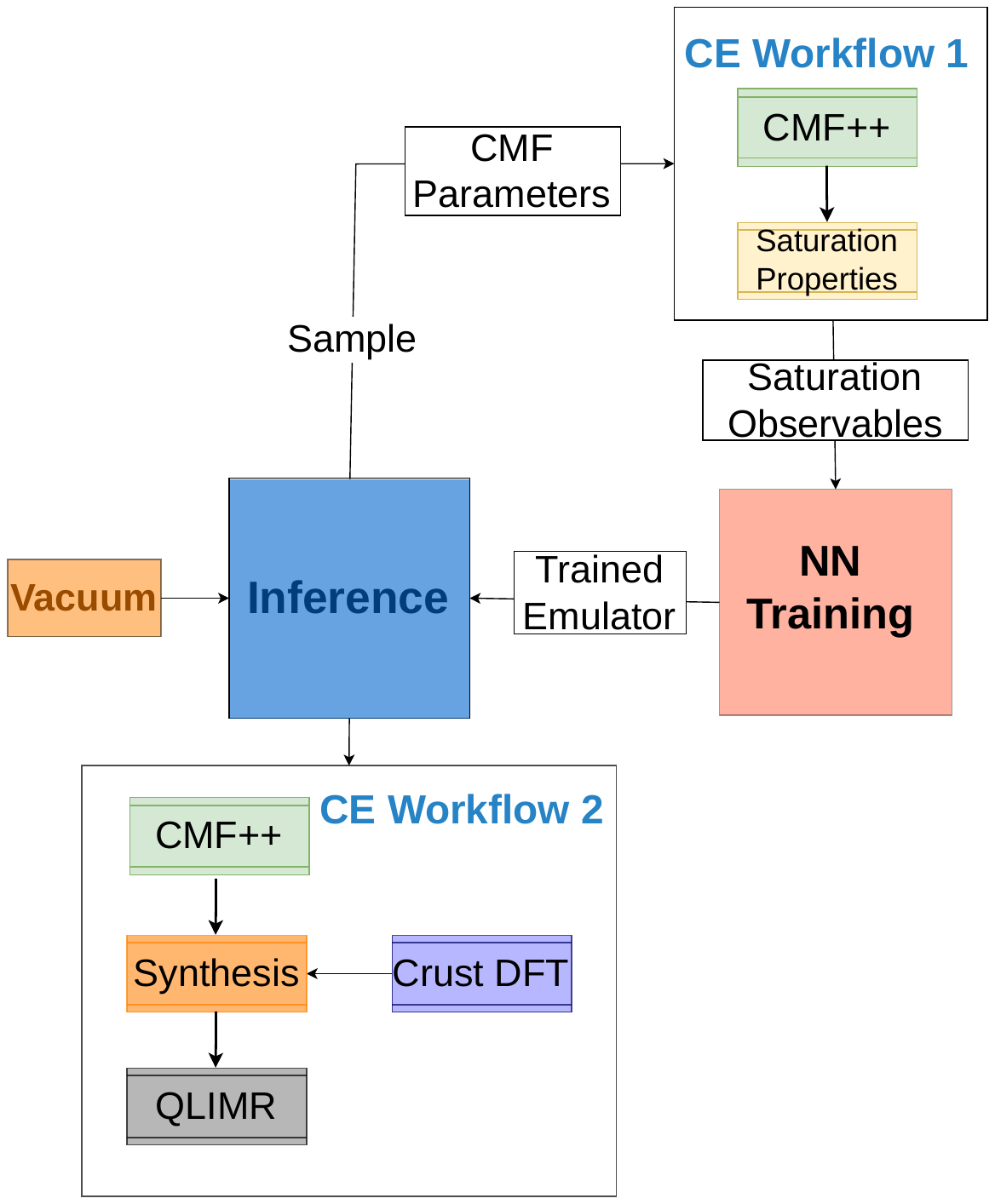}
    \caption{Schematic diagram of the Calculation Engine workflows, emulation and inference pipeline we use.  Our inference is performed as a component of the loop at the top of the diagram.  This loop consists of solving for the saturation properties of proposed \cmf{} parameters, emulating the mapping of \cmf{} parameters to saturation observables, and finally performing the full Bayesian inference using the emulator (as well as the vacuum observables, which do not require an emulator). From the inference, new proposed \cmf{} parameter sets can be drawn, and the loop can be repeated. The main loop is upstream of solving for macroscopic neutron star properties which occurs in the bottom leg of the diagram.  }.
    \label{fig:workflow}
\end{figure}

We collect here the software stack and sampler settings used in the inference. The \cmf{} evaluations used to generate emulator training data are performed with the open-source \cmfpp equation-of-state module \cite{zenodo_cmf}. Each configuration is evaluated on a dense grid of baryon and charge chemical potentials near saturation density, with spacing of $0.1\,\rm{MeV}$ in $\mu_B$  and $1\, \rm{MeV}$ in $\mu_Q$.  The saturation and symmetry properties are evaluated using the {\sc Saturation Properties} module \cite{pelicer_2026_21079718}, which will be released with this paper and made available through the MUSES {\sc Calculation Engine}~\cite{zenodo_cmf,calculation_engine_all_versions,manning_2025_14721912,ReinkePelicer:2025vuh}.  For each choice of \cmf{} parameters, \cmfpp is used to compute nucleonic infinite-matter solutions on the grid needed to extract the saturation observables.  We do not include hyperons or quarks in these saturation-property calculations.  Besides the parameters varied in the inference, we use the standard settings described in the \cmfpp documentation.  Configuration files and scripts needed to reproduce the results will be made publicly available.

The neural-network emulator is implemented in {\sc pytorch}, as described in \Cref{sec:neural-network-details}.  The emulator is used only to accelerate the map from \cmf{} parameters to saturation observables inside the sampler; the training and validation data are generated from direct \cmfpp calculations.

For stochastic sampling, we use the nested sampler {\sc dynesty}~\cite{Speagle20}, with extensions implemented in the {\sc bilby} inference library~\cite{Ashton2019}.  See Ref.~\cite{Skilling2006} for an introduction to nested sampling.  We use 1000 live points and the \texttt{act-walk} Markov-chain Monte Carlo method for generating new live points, as implemented in {\sc bilby}.  We adopt a stopping criterion $d\log\mathcal Z=0.4$, which is weaker than would be appropriate for precision evidence calculations.  This is sufficient for our purposes because the evidence is used only as a rough diagnostic, while uncertainties associated with the priors, likelihood construction, and emulator approximation dominate the numerical evidence error.  We have checked that tightening this criterion does not substantially change the final results of the iterative scheme.

In \Cref{fig:workflow}, we show a schematic overview of the inference pipeline for the \cmf{} parameters and how they connect to the Calculation Engine workflows. The first workflow maps sampled \cmf{} parameters through \cmfpp and the Saturation Properties module to the saturation observables used to train the neural-network emulator. The trained emulator is
then used in the inference loop, using the posterior of the vacuum parameters. Posterior samples are then propagated through a second Calculation Engine workflow, which computes the beta-equilibrated \cmf{} and crust-DFT EoS, matches them through the synthesis module, and calculates neutron star properties with the QLIMR module.

\section{Bayesian exploration of CMF parameter space}
\label{sec:results}

We now discuss the results of our inference. We start in \Cref{sec:vacuum-results} with the parameters $\theta_{\rm vac}$ that can be highly constrained based on measurements of the vacuum masses of nucleons and hyperons.  In \Cref{sec:nuclear-matter-results}, we discuss the full inference of parameters that influence the properties of nuclear matter near saturation density, $\theta_{\rm med}$.  

\subsection{Vacuum}
\label{sec:vacuum-results}
\begin{figure}
    \centering
    \includegraphics[width=0.99\linewidth]{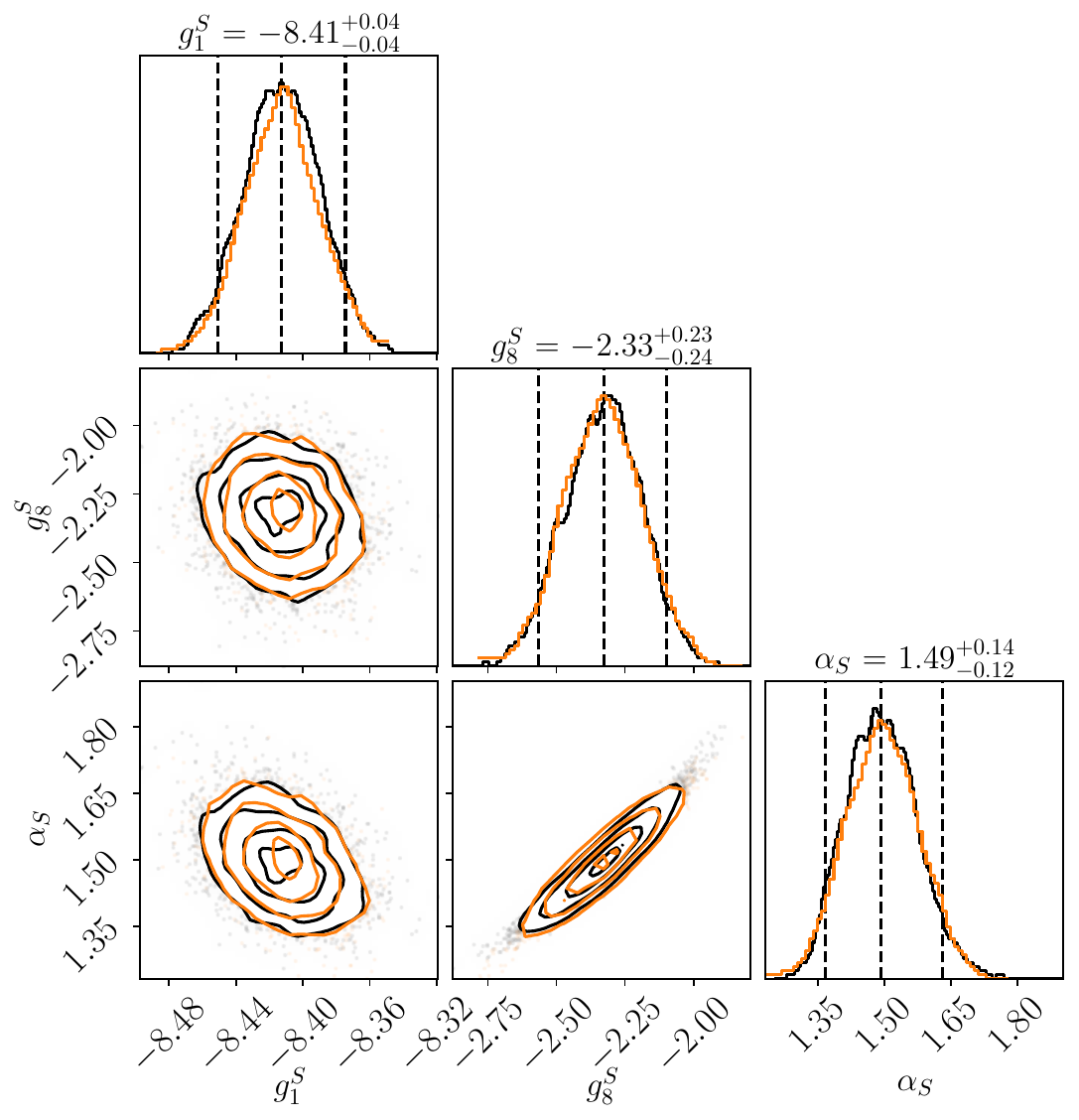}
    \caption{Posterior distribution for the scalar-sector parameters constrained by vacuum baryon masses. The black contours show the sampled posterior, and the orange contours show the multivariate Gaussian approximation used as the prior for the in-medium inference.  Level sets of the posterior density are shown corresponding to $[0.118, 0.393, 0.675, 0.864]$ of the maximum.  Priors for these parameters are given in \Cref{tab:priors-no-m-zero}.  }
    \label{fig:vacuum-parameters-corner}
\end{figure}

We display the inferred values for the parameters, which are highly constrained by vacuum hadron masses in \Cref{fig:vacuum-parameters-corner}.  For this analysis, priors are uniform, and the likelihood is very nearly Gaussian, so the posterior is very nearly Gaussian.  We therefore fit a multivariate Gaussian distribution to the posterior for use in the sampling using nuclear properties (\Cref{sec:nuclear-matter-results}).  The assumed systematic uncertainty in modeling the hadron masses dominates the uncertainty on the parameters.  By far, the most uncertain parameter in a fractional sense is $\alpha_S$ =\alphasnomzerovalue{}. 
In the interaction term of \Cref{eq:L_BM}, $\alpha_S$ controls the relative weight of the antisymmetric and symmetric interactions between the octet sector of the scalar mesons and the baryons: $\alpha_S=1$ and $\alpha_S=0$ correspond to the cases in which only the antisymmetric or symmetric terms contribute, respectively, while values outside this interval change the relative sign of these contributions with respect to $g_8^S$.
Meanwhile, $g_1^S$, which controls the baryon octet interaction with the singlet sector of the scalar mesons, is highly constrained, with fractional precision $\result{\sim 0.5\%}$. We find that this parameter is critically related to our choice to fix $m_0=0$, and we discuss alternatives in \Cref{app:mzero-nonzero}.  The parameter $g_8^S$ is constrained at an intermediate level, due to an apparent correlation with the poorly constrained parameter $\alpha_S$; $g_{8}^{S}$ is constrained to lie within  $\result{\sim 10\%}$ of its inferred value.   

\subsection{In-medium}
\label{sec:nuclear-matter-results}

We now discuss our results for properties of matter defined at saturation density.  Computing these properties requires solving the \cmf{} equations of motion in-medium; in particular, we do that on a structured grid of baryon chemical potentials and charge chemical potentials, which is equivalent to an unstructured grid of baryon densities and charge fractions. As discussed before, while the computational time is small for a few parameter sets, the increased computational cost of calculating properties of infinite matter over millions of parameter sets is costly. An emulator accelerates the process substantially.  
In \Cref{sec:emulator-performance}, we discuss the performance of the emulator within the inference problem.  In \Cref{sec:in-medium-inference-results}, we apply the emulator to the inference problem.  

\subsubsection{Performance of the emulator}
\label{sec:emulator-performance}

We first discuss the performance of the emulator introduced in \Cref{sec:emulator-and-iterative-scheme} during practical application within our iterative scheme.  In  general, we find that emulation of nuclear properties is possible at a comparable level of precision to experimental constraints, though ultimately the density of training data throughout parameter space determines the effectiveness of the emulator.  The effectiveness of the emulator with even a limited amount of training data is potentially surprising, and may point to structure in the space of plausible configurations, which we explore in detail in \Cref{sec:in-medium-inference-results}.

\begin{figure*}
    \centering
    \includegraphics[width=0.49\linewidth]{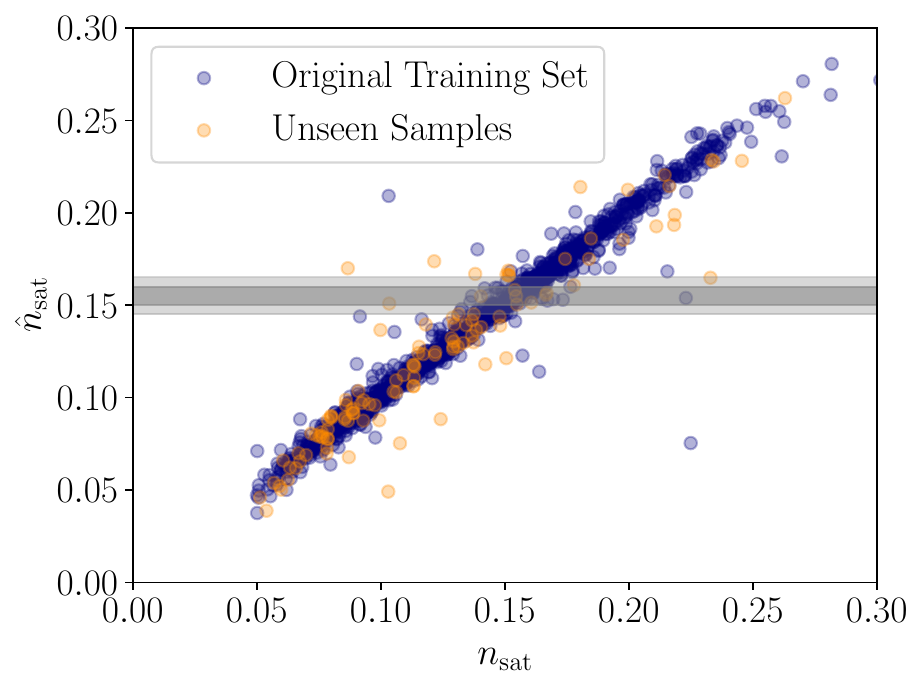}
    \includegraphics[width=0.49\linewidth]{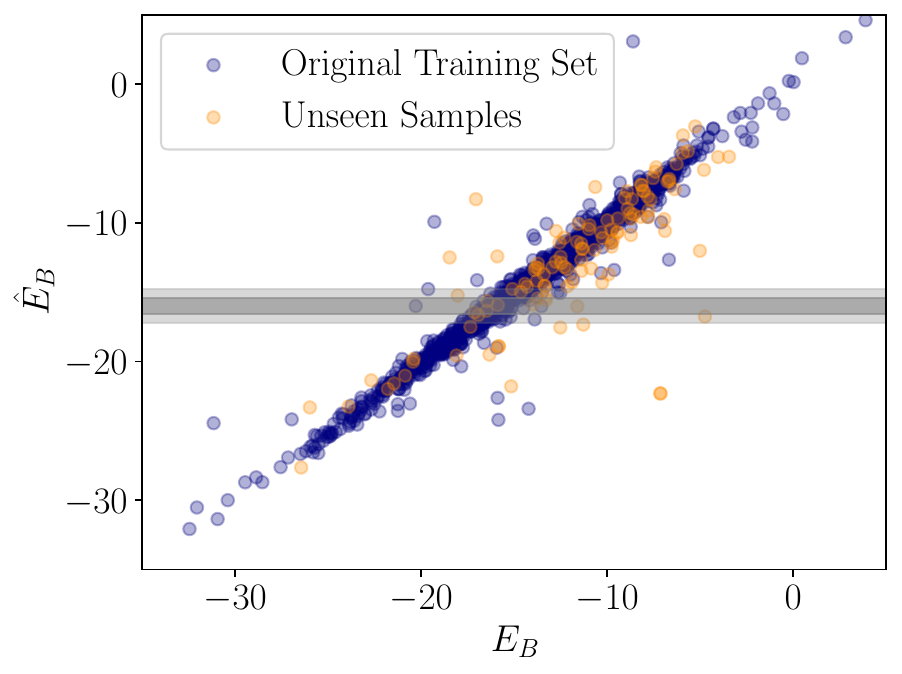}
    \includegraphics[width=0.49\linewidth]{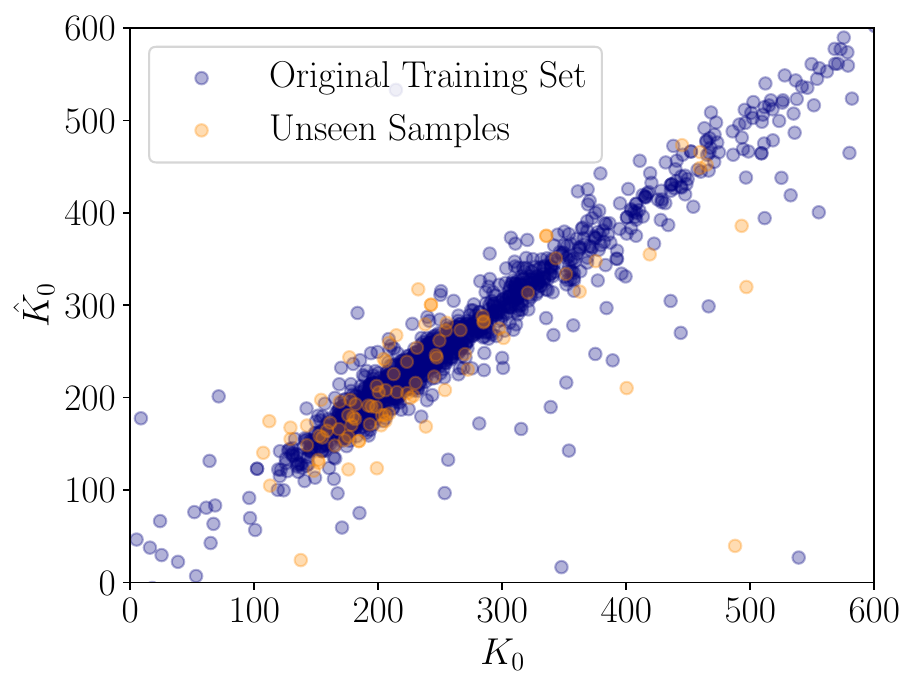}
    \caption{Emulator performance for symmetric nuclear matter observables.   Truth is shown on the x-axis and the emulated results are shown with a hat on the y-axis. Blue points show samples from the original training set, while orange points show samples generated during iteration 8 that were not used for emulator training.  In dark and light gray horizontal bars we display the chosen 1- and 2-sigma regions for $\nsat$ and $\Esat$. We do not show the analogous bars for $K$ because the width would be a large fraction of the entire plot (200 and 400 respectively).}
    \label{fig:emulator-validation-symmetric}
\end{figure*}

\begin{figure}
    \centering
    \includegraphics[width=0.99\linewidth]{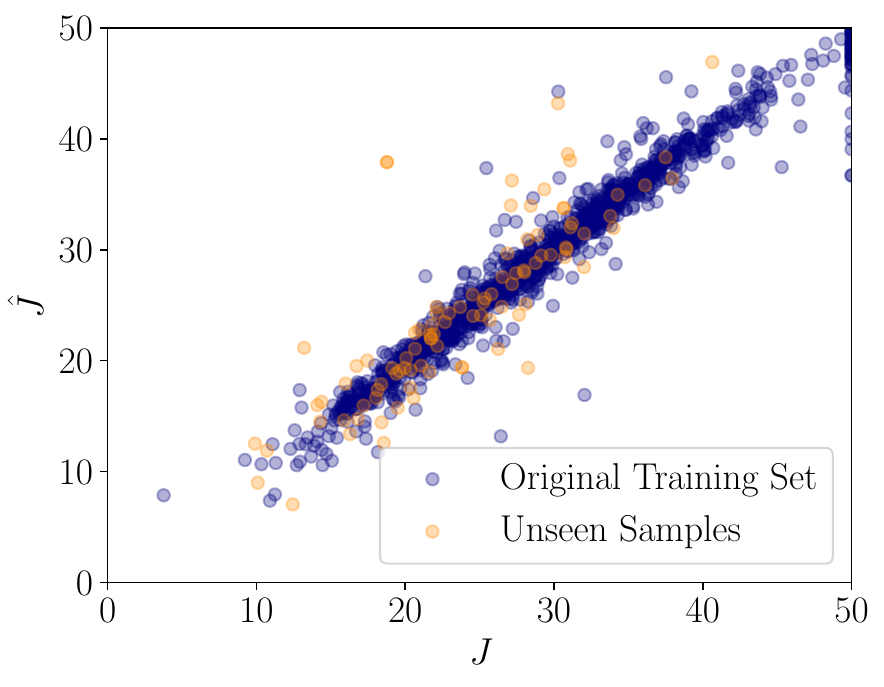}
    \includegraphics[width=0.99\linewidth]{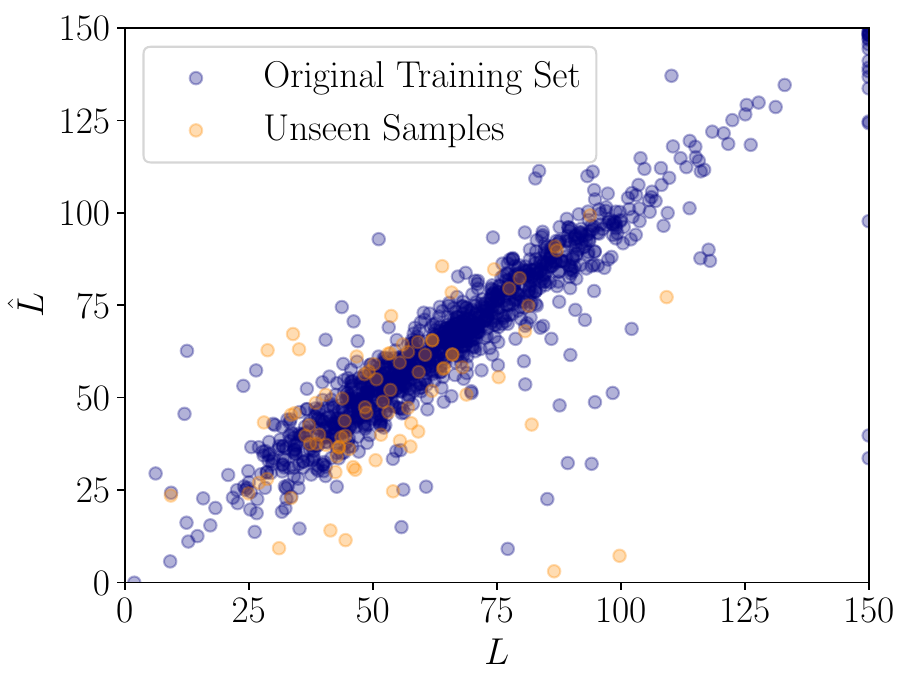}
    \caption{Emulator performance for the symmetry energy expansion parameters.  Blue points show the emulator performance on the original training set after iteration 8. Orange points show the emulator performance on samples that were generated during iteration 8 but that were not used for training.}
    \label{fig:emulator-validation-asymmetric}
\end{figure}

We first give a representative picture of the emulator performance in \Cref{fig:emulator-validation-symmetric}, focusing on the properties of isospin-symmetric ($\mu_Q=0$) nuclear matter at saturation density,  and in \Cref{fig:emulator-validation-asymmetric} focusing on the isospin-asymmetric parameters $\Esym$ and $\Lsym$, which requires calculating the 2-dimensional EoS table in $\mu_B$, $\mu_Q$.
Unless otherwise stated, we show results for the \emph{second} pass of the iterative scheme, for which the emulator is more accurate. In this comparison, the direct \cmfpp calculation is shown on the horizontal axis and the emulator prediction on the vertical axis. We denote emulator predictions with a hat, e.g. $\hat n_{\sat}$, to distinguish them from the saturation properties extracted directly from solving the equations. Blue points correspond to samples from the original training set, while orange points correspond to samples generated during the iterative procedure that were not used to train the emulator.

The agreement is generally good, although some individual samples are still mispredicted. As expected, the unseen samples show somewhat more scatter than the training samples. For $\nsat$ and $\Esat$, we also show the adopted $1$-- and $2$--$\sigma$ uncertainty bands, which indicate that for most samples the emulator error is smaller than, or comparable to, the uncertainty scale used in the likelihood. Although we do not perform this check for every posterior sample in the present work, one could also recompute the saturation properties directly for the best EoS candidates selected by the emulator. This would still be far cheaper than using the full \cmfpp calculation inside the nested sampler.

A single summary statistic for emulator performance is not unique, because the answer depends both on which region of parameter space is used for the comparison and how errors in different observables are compared. Here, we quantify the error through the distribution of $X-\hat X$ for each saturation observable $X$. In \Cref{tab:errors-seen}, we show the median and symmetric $68\%$ credible interval for this error on the original training data. In \Cref{tab:errors-unseen}, we show the same quantity on new samples generated during the iterative procedure, following \Cref{sec:emulator-and-iterative-scheme}. We use the \emph{first} pass of the algorithm for this comparison as a stress test, since that pass intentionally starts with limited training data. The errors on the original training set do not improve dramatically with iteration, which we attribute to two effects. First, the original training set becomes a progressively smaller fraction of the full training set as new samples are added. Second, many of the original samples were drawn from broad priors and are therefore highly unphysical, with saturation properties orders of magnitude away from the empirical region; these samples are strongly affected by the clamping procedure discussed in \Cref{sec:neural-network-details}.

On the other hand, the emulator performance notably improves as new training data is incorporated, when testing it against \emph{unseen} data.  This is likely because the bulk of new samples, under our scheme, falls near the existing posterior distribution, which is stable from iteration to iteration.  Therefore, as further iterations are performed, we obtain more training samples exactly in the regions where future iterations will be sampling.  We discuss this further in \Cref{app:stress-tests}. 

By iteration 11, the median emulator bias is small and the typical errors are comparable to the adopted saturation-property uncertainties. This supports using the emulator for exploratory parameter-space mapping, although the posterior should still be interpreted as conditional on the emulator approximation.
As we discuss below, however, we do not find that the posterior has stabilized by iteration 11.  We suspect that as long as the prior on the \cmf{} parameters is increased, the posterior will continue to grow.  Therefore, the choice of stopping iteration is somewhat arbitrary.

\begin{table*}[]
\centering
\begin{tabular}{c|ccccc}
\hline
Iteration & $\nsat$ & $\Esat$ & $\Ksat$ & $\Esym$ & $\Lsym$ \\
\hline
0 & $-0.000^{+0.014}_{-0.019}$ & $0.000^{+1.441}_{-1.676}$ & $-0.032^{+268.820}_{-585.062}$ & $-0.012^{+25.774}_{-13.752}$ & $-2.102^{+1997.809}_{-2692.671}$ \\
1 & $-0.000^{+0.012}_{-0.017}$ & $0.012^{+1.629}_{-2.472}$ & $-0.912^{+60.031}_{-115.986}$ & $-0.023^{+4.706}_{-35.542}$ & $-0.176^{+1021.314}_{-45.875}$ \\
2 & $0.000^{+0.022}_{-0.008}$ & $0.018^{+0.698}_{-0.672}$ & $0.912^{+61.651}_{-79.090}$ & $0.046^{+1.362}_{-28.615}$ & $-0.113^{+1020.946}_{-39.481}$ \\
3 & $0.000^{+0.024}_{-0.005}$ & $0.006^{+0.242}_{-2.885}$ & $-0.773^{+436.306}_{-14.784}$ & $0.153^{+1.249}_{-50.361}$ & $-0.258^{+1092.237}_{-12.073}$ \\
4 & $-0.001^{+0.035}_{-0.004}$ & $0.009^{+0.378}_{-1.622}$ & $-7.556^{+75.339}_{-93.919}$ & $0.030^{+2.861}_{-29.121}$ & $-1.887^{+1020.888}_{-61.287}$ \\
5 & $-0.000^{+0.006}_{-0.008}$ & $-0.066^{+0.394}_{-1.393}$ & $-3.590^{+228.958}_{-29.298}$ & $-0.065^{+1.450}_{-28.217}$ & $1.845^{+1026.189}_{-51.312}$ \\
6 & $-0.001^{+0.019}_{-0.012}$ & $0.050^{+1.047}_{-1.794}$ & $-1.611^{+512.103}_{-37.645}$ & $-0.425^{+2.614}_{-51.208}$ & $1.141^{+1079.321}_{-52.291}$ \\
7 & $0.001^{+0.024}_{-0.014}$ & $-0.234^{+1.969}_{-2.205}$ & $-7.989^{+161.947}_{-106.889}$ & $0.617^{+3.943}_{-48.004}$ & $0.711^{+1028.738}_{-67.568}$ \\
8 & $-0.002^{+0.006}_{-0.006}$ & $0.154^{+1.000}_{-1.242}$ & $4.099^{+312.696}_{-29.357}$ & $-0.545^{+1.904}_{-50.706}$ & $-0.347^{+1029.850}_{-12.170}$ \\
9 & $-0.000^{+0.027}_{-0.006}$ & $-0.045^{+0.445}_{-0.896}$ & $8.213^{+79.569}_{-54.985}$ & $-0.199^{+1.810}_{-31.886}$ & $1.197^{+1020.260}_{-44.396}$ \\
10 & $-0.001^{+0.004}_{-0.024}$ & $-0.002^{+0.451}_{-1.338}$ & $4.563^{+130.226}_{-96.630}$ & $-0.141^{+4.853}_{-41.069}$ & $0.215^{+1020.125}_{-80.903}$ \\
11 & $0.001^{+0.013}_{-0.007}$ & $-0.161^{+0.733}_{-1.131}$ & $11.139^{+72.113}_{-26.291}$ & $-0.116^{+2.268}_{-28.264}$ & $1.410^{+1028.107}_{-11.163}$ \\
\hline
\end{tabular}
\caption{Median and $68\%$ credible region boundaries on the emulator errors on data that \emph{has} been included in its training set.  Columns represent the error in the given observable for the given data using  the emulator from that iteration.  Therefore the emulator used is what is changing from row to row.  The anomalously large error on $L$ is due to our clipping procedure on input data, which causes us to intentionally mispredict the saturation properties of samples if they are very far from physical values.  Values are shown for 12 iterations of the algorithm presented in \Cref{sec:iterative-emulation}.}
\label{tab:errors-seen}
\end{table*}

\begin{table*}[]
\centering
\begin{tabular}{c|ccccc}
\hline
Iteration & $\nsat$ & $\Esat$ & $\Ksat$ & $\Esym$ & $\Lsym$ \\
\hline
0 & $0.009^{+0.024}_{-0.020}$ & $-0.288^{+2.479}_{-2.584}$ & $-59.277^{+210.948}_{-311.290}$ & $3.715^{+17.790}_{-12.717}$ & $-76.803^{+1099.192}_{-1326.796}$ \\
1 & $0.001^{+0.019}_{-0.016}$ & $0.122^{+1.767}_{-1.663}$ & $12.868^{+125.783}_{-81.752}$ & $1.046^{+7.360}_{-4.551}$ & $0.207^{+32.267}_{-31.998}$ \\
2 & $0.001^{+0.019}_{-0.013}$ & $0.077^{+1.176}_{-1.539}$ & $12.511^{+132.548}_{-118.783}$ & $0.406^{+7.713}_{-6.010}$ & $3.247^{+35.746}_{-36.817}$ \\
3 & $0.003^{+0.012}_{-0.009}$ & $0.078^{+0.677}_{-0.854}$ & $9.607^{+123.326}_{-63.524}$ & $0.695^{+3.208}_{-4.235}$ & $6.024^{+26.573}_{-23.772}$ \\
4 & $-0.002^{+0.012}_{-0.012}$ & $0.173^{+1.201}_{-1.206}$ & $7.287^{+130.293}_{-91.830}$ & $-1.064^{+5.216}_{-6.802}$ & $-1.436^{+33.880}_{-30.615}$ \\
5 & $-0.001^{+0.013}_{-0.013}$ & $-0.072^{+1.298}_{-1.175}$ & $17.114^{+84.871}_{-75.586}$ & $-0.342^{+3.514}_{-5.302}$ & $3.351^{+26.651}_{-28.068}$ \\
6 & $-0.001^{+0.010}_{-0.012}$ & $0.294^{+1.543}_{-1.425}$ & $-15.535^{+100.843}_{-83.895}$ & $-0.433^{+4.043}_{-6.414}$ & $-1.531^{+20.075}_{-538.366}$ \\
7 & $-0.001^{+0.012}_{-0.011}$ & $-0.021^{+1.287}_{-1.521}$ & $-10.212^{+82.576}_{-81.402}$ & $0.698^{+5.327}_{-4.476}$ & $-5.049^{+19.090}_{-28.072}$ \\
8 & $-0.001^{+0.012}_{-0.013}$ & $0.098^{+1.083}_{-1.600}$ & $15.377^{+97.167}_{-84.437}$ & $-0.403^{+5.703}_{-3.817}$ & $0.468^{+21.382}_{-119.430}$ \\
9 & $-0.002^{+0.014}_{-0.011}$ & $0.068^{+1.640}_{-1.495}$ & $14.729^{+118.944}_{-68.790}$ & $0.046^{+3.848}_{-4.730}$ & $-1.064^{+18.450}_{-31.630}$ \\
10 & $-0.003^{+0.008}_{-0.018}$ & $0.069^{+1.285}_{-0.976}$ & $6.019^{+86.182}_{-55.769}$ & $0.153^{+2.067}_{-5.612}$ & $-6.133^{+16.993}_{-541.586}$ \\
11 & $-0.000^{+0.013}_{-0.008}$ & $-0.017^{+1.352}_{-1.378}$ & $7.526^{+97.831}_{-35.084}$ & $-0.348^{+2.028}_{-3.234}$ & $0.344^{+20.113}_{-25.802}$ \\
\hline
\end{tabular}
\caption{Median and $68\%$ credible region boundaries on the emulator errors on data that has \emph{not} been included in its training set. Otherwise, descriptions and caveats are the same as \Cref{tab:errors-seen}. }
\label{tab:errors-unseen}
\end{table*}

Nonetheless, the emulator performance is precise compared to the range of possible values of saturation parameters.  We find that sampling at random from the prior on \cmf{} parameters, only $\sim \mathrm{a\ few}$ in hundreds of samples are close to having physical saturation parameters (many of the EoSs do not saturate).  On the other hand, among parameter combinations inferred by the emulator to have nearly physical parameters, almost all of them ($\sim 95\%$) display nearly physical saturation when their saturation properties are computed directly.   Additionally, we find that the emulator errors are not strongly correlated between parameters (Pearson coefficient $r^2 < 0.5$ ).  Therefore, we proceed to interpret the parameter inference results, recognizing that inferred posteriors represent the posteriors on parameters \emph{under the approximation of the model being provided by the emulator}.  We discuss in \Cref{sec:discussion} how exactly the systematic uncertainty introduced by using this choice compares to other sources of systematic uncertainty.

\subsubsection{Inference results}
\label{sec:in-medium-inference-results}
\begin{figure}
    \centering
    \includegraphics[width=0.99\linewidth]{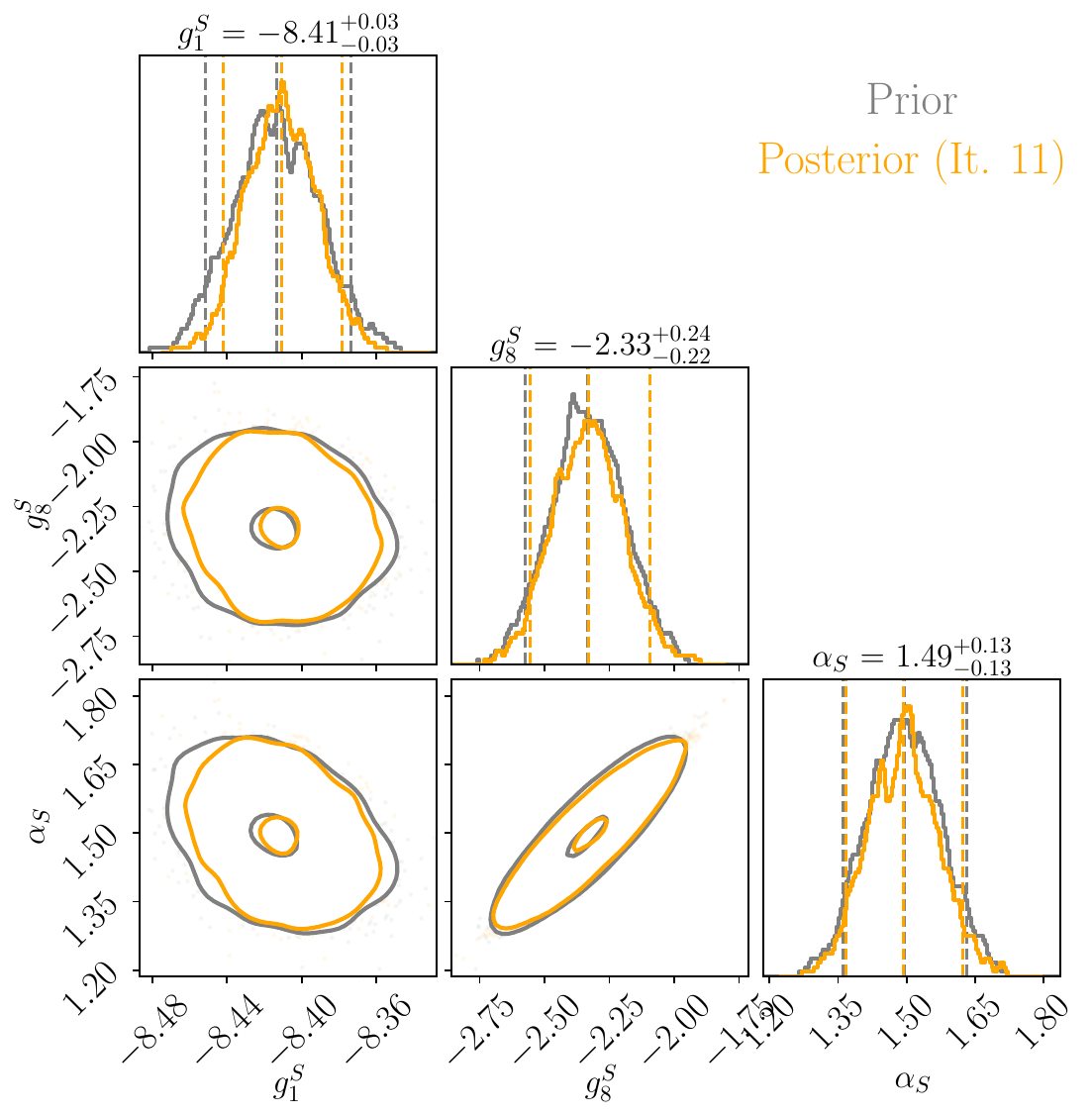}
    \caption{In gray, the inference of \Cref{sec:vacuum-results} for the parameters $g_1^S, g_8^S$, and $\alpha_S$, which becomes the prior distribution for inference of parameters relevant for nuclear matter at saturation.  In gold, the posterior after iteration 11 on the same parameters. Contours denote approximate boundaries of 50\% and 90\% credible regions. }
    \label{fig:scalar-nucleon-in-medium-inference}
\end{figure}

We now discuss the inferred posterior on all of the relevant \cmf{} parameters for the nuclear model under the inference framework laid out in \Cref{sec:bayesian_analysis}. To begin with, we discuss the parameters that have already been inferred under the first hierarchical step in \Cref{sec:vacuum-results}. We show the inferred posterior on the vacuum parameters after iteration 11 in \Cref{fig:scalar-nucleon-in-medium-inference}, versus the initial prior used for the initial iteration (iteration 0). We find the posterior distributions on these parameters are effectively identical to their distribution using the hadron-mass data alone (which is here used as the prior), indicating that the additional likelihood of inferring saturation properties is only a small perturbation to the constraints provided by vacuum mass measurements (even with our conservative estimates of systematic uncertainty). 

\begin{figure*}
    \centering
    \includegraphics[width=0.99\linewidth]{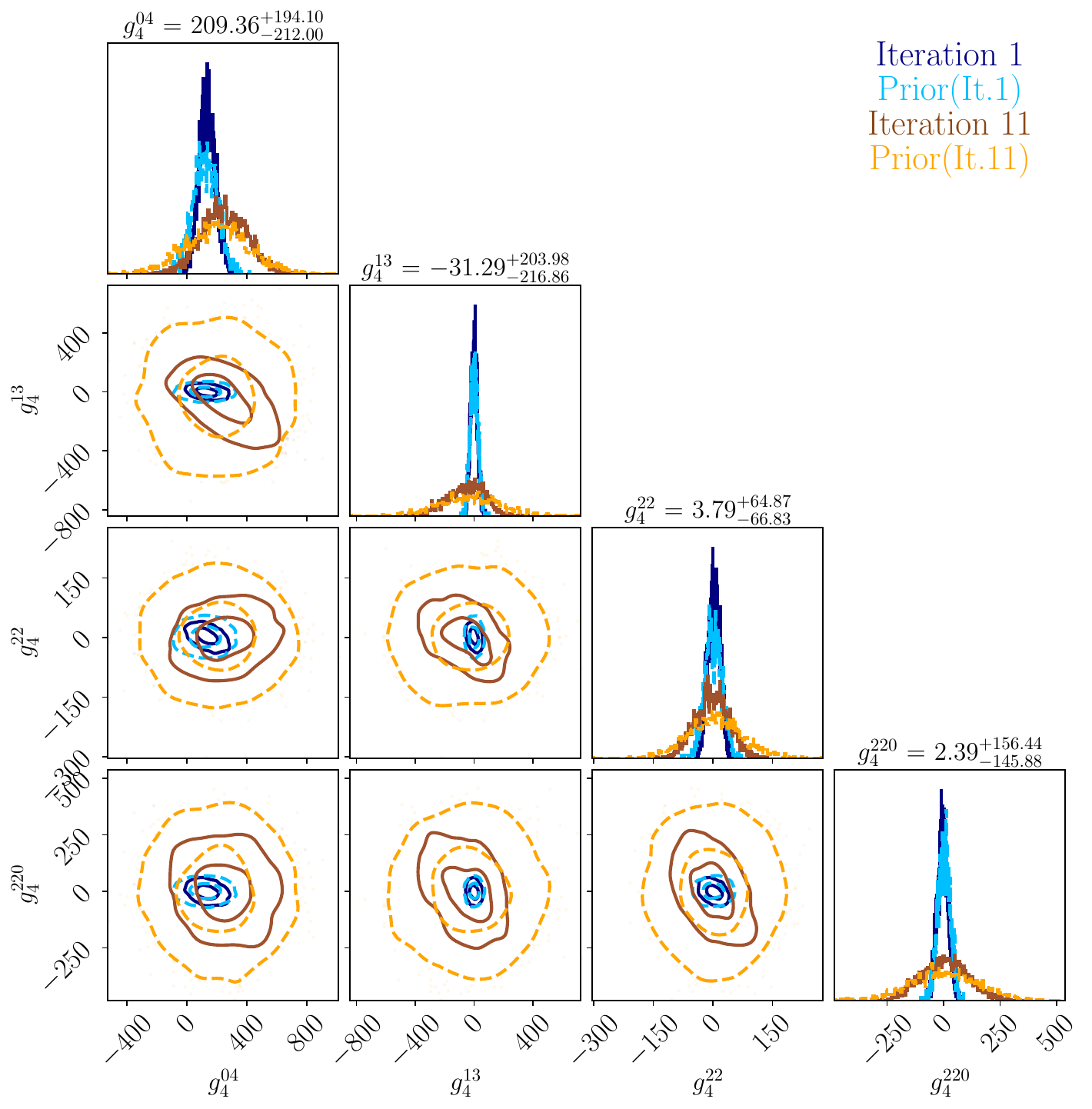}
    \caption{Posterior on the vector couplings after the first update (solid blue)  versus for the final iteration (solid brown).  The corresponding prior distributions are shown in dashed light-blue and dashed gold lines  for iterations 1 and 11 respectively. Contours mark the boundaries of approximate 50\% and 95\% credible regions.   }
    \label{fig:vector-self-coupling}
\end{figure*}

\begin{figure}
    \centering
    \includegraphics[width=0.99\linewidth]{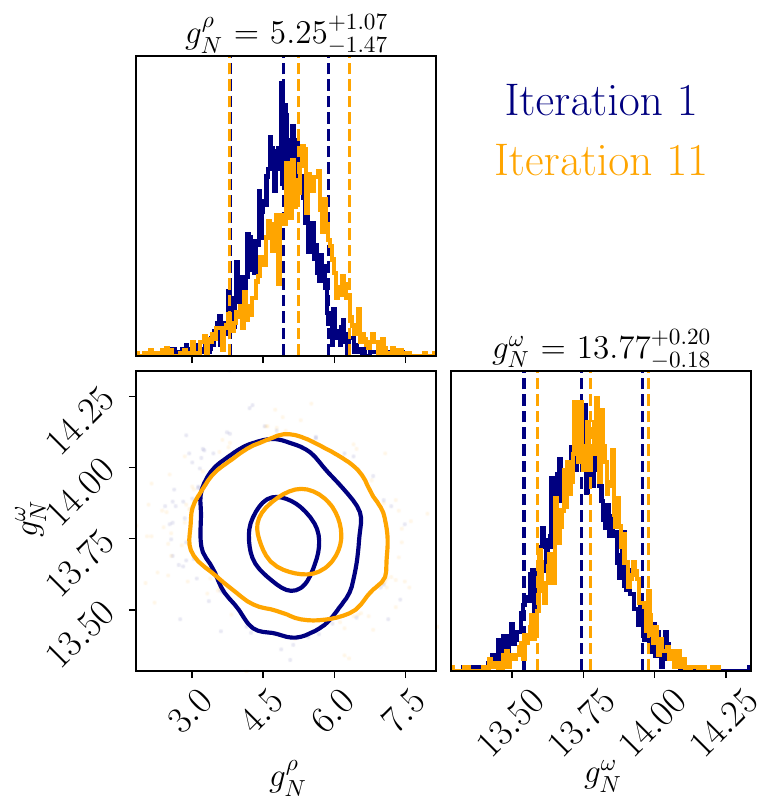}
    \caption{Posterior on the vector nucleon couplings after iteration 1 (blue) and the final iteration (gold).}
    \label{fig:vector-nucleon-coupling}
\end{figure}

Moving on, we examine the inferred values of the vector self-couplings of the \cmf{} model.  In \Cref{fig:vector-self-coupling}, we display samples of the four vector self coupling terms from the \cmf{} model after iteration 1 and iteration 11 in blue and brown respectively, and the vector to nucleon couplings in \Cref{fig:vector-nucleon-coupling}. Priors are marked with lighter shades and dashed lines. We also show both the $50\%$ and $95\%$ credible regions there.  We see that, while the posterior expands significantly from iteration 1 to 11, the ratio of the width of the posterior to the width of the prior changes much less.  Therefore, broadly speaking, it appears as if the inference is \emph{prior driven} for the self-couplings. 

However, the likelihood is nonetheless very informative.  In the full 9-dimensional parameter space, only a small fraction of samples are plausible, given the likelihood we choose.  For example, under one choice of random seed, sampling $10^7$ points from the prior yields only $981$ points with likelihood values greater than $10^{-5} \times \mathcal L_{\rm max}$ (the maximum likelihood). The reason the prior and posterior appear to coincide so closely is simply that the structure of the 9-dimensional likelihood surface is, for the most part, completely scrambled by the process of marginalization.  We present a toy model of a 2-dimensional inference problem that displays the same characteristics in \Cref{app:toy-model}.

A natural question, however, is whether there is a higher dimensional structure that cannot be probed from simple 2-dimensional projections.  We find no evidence of nontrivial large-scale structure. We discuss this in more detail in \Cref{app:data-structure}, and here we briefly summarize our findings.  First, using a $k$-means clustering algorithm implemented in {\sc scikit-learn}~\cite{scikit-learn}, and trying to identify the preferred number of clusters using the elbow method~\cite{Thorndike_1953}, we find no obvious indication of a preferred number of clusters.  
Rather, increasing the number of clusters up to $k \gtrsim 100$ continues to improve the explanatory power, while leaving the silhouette score nearly unchanged.

We further attempt to identify nontrivial topological structure by computing the persistent homology~\cite{Edelsbrunner2002, Zomorodian2005} (see Ref.~\cite{JodelleKemme:2025} for a pedagogical introduction and contemporary references) of the posterior distribution using {\sc ripser}~\cite{Bauer2021Ripser}. Persistent homology is designed to capture ``important"  features in the distribution of point cloud data, i.e.~those that are unlikely to be due to sampling noise.  Equivalently, persistent homology captures topological features that are present when the structure is probed at a diverse range of length scales. We compute the lowest dimensional persistent homology groups $H_0, H_1$ and $H_2$ and find the resulting diagrams are qualitatively very similar to the diagrams produced from a multidimensional Gaussian distribution.  We further compute $H_4$, and $H_5$ at lower sampling resolution to reduce computational cost, and again find no obvious important features.  This indicates that at least up to dimension $d=5$, there is likely no substantial large-scale structure in the distribution in the form of e.g. ``islands" ($H_0$) ``loops" ($H_1$) or ``voids" ($H_2$). See ~\Cref{app:data-structure} for more details.

\begin{figure}
    \centering
    \includegraphics[width=0.99\linewidth]{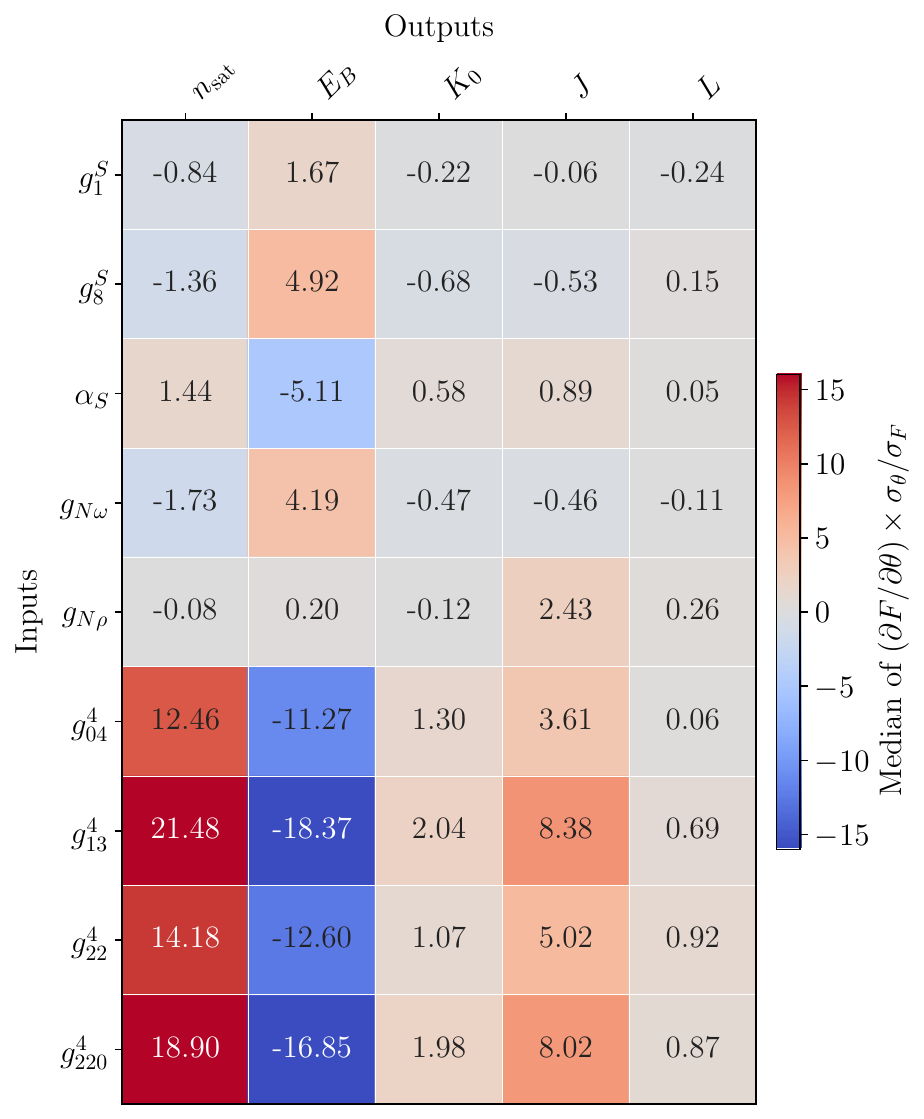}
    \caption{The median value of $\partial  F_i/\partial  \theta_j$ $\sigma_{\theta_j}/\sigma_{F_i}$ for observables  $f$ (where $\sigma_{F_i}$ is given by \Cref{tab:mass-measurements}), and $\sigma_{\theta_j}$ is measured from the posterior (e.g. \Cref{fig:vector-self-coupling,fig:vector-nucleon-coupling}). To compute  $\partial  F_i/\partial  \theta_j$ $\sigma_{\theta_j}/\sigma_{F_i}$, we use the emulator and the posterior from the final iteration of the algorithm (iteration 11).  }
    \label{fig:autodiff-dependency}
\end{figure}

Despite the lack of large scale structure in the posterior, we do find evidence of structure at much \emph{smaller} scales. 
We produce local estimates of the scale of variation of the likelihood function by computing the change of the observables $F_{\rm sat}$, with respect to the relevant \cmf{} parameters. In \Cref{fig:autodiff-dependency}, we display estimates for the median value of $(\partial  F_i / \partial  \theta_j) \sigma_{\theta_j} / \sigma_{F_i}$, which measures how much, relative to its allowed range,  each observable ($F_i$) varies with respect to each input ($\theta_i$) over the plausible range of that input.
The general interpretation of this number is as follows: 
\begin{itemize}
    \item $(\partial  F_i / \partial  \theta_j) \sigma_{\theta_j} / \sigma_{F_i} \ll 1$: The observable has very little dependence, or no average dependence on the particular parameter.  
    \item $(\partial  F_i / \partial  \theta_j) \sigma_{\theta_j} / \sigma_{F_i} \approx 1$: The observable depends on this parameter, and the width of the posterior on this parameter is likely driven by the uncertainty on the given observable.
    \item $(\partial  F_i / \partial  \theta_j) \sigma_{\theta_j} / \sigma_{F_i} \gg 1$: The observable depends on this parameter, but the width of the posterior on the given parameter is much larger than would be predicted via na\"ive uncertainty propagation.  Degeneracies between other parameters and this parameter are likely relevant.  
\end{itemize}

Let us focus first on the dependence of $\nsat$ and $\Esat$ on $\alpha_S$ and the vector-self couplings. The fact that these dependences exceed $\sim 1$ implies that extrapolating local gradients of $\nsat$ and $\Esat$ to the full range of the posterior on the parameters would predict $\gg 1 \sigma$ changes in the observables.  Therefore, for these parameters, it is clear that local variations in the observables do not translate directly to global changes in the observables, i.e. the mapping from inputs to outputs is highly nonlinear over the allowed parameter space.  

For the other relations, $(\partial F_i / \partial  \theta_j) \sigma_{\theta_j} / \sigma_{F_i} < 1$ may mean that the output is mostly independent of that input, or that the median value of this dependence over the posterior is coincidentally close to zero.  In general, the likelihood changes on a scale that is much smaller than the global characteristic scale of the posterior (set by $\sigma_{\theta_i}$, for parameters $\theta_i$), which implies that there is substantial structure in the likelihood that is not visible in the $1$---and $2$---dimensional marginal distributions. For example, because the entry for $g^4_{13}$--$\nsat$ is $21.48$, the   rate of change of the likelihood in the $g^4_{13}$ direction is about  $21.48$ times the global rate of change in the posterior for $g^4_{13}$. This means the likelihood must have small scale structure.

\subsection{The nuclear matter EoS and neutron star properties}

Since our inference considers only properties defined for matter near saturation density, hyperons, quarks, or other degrees of freedom are not relevant at the level of the uncertainty of our posteriors.  We now evaluate the \cmf{} model more broadly to compute the nuclear EoS at large densities and large isospin asymmetries characteristic of neutron star matter.  The entire system is evaluated in $\beta$-equilibrium (with only electrons) to ensure charge neutrality using the {\sc Lepton} module of the Calculation Engine~\cite{calculation_engine_all_versions}.  At densities below the regime of applicability of the \cmf{} model, we use a precomputed crust model using the {\sc crust-DFT} module, containing a statistical ensemble of nuclei~\cite{Steiner:2012rk,Steiner:2014pda,steiner_2025_14714273}.  We stitch the crust and core solutions using the {\sc synthesis module}, choosing a stitching density of $n_B=0.08\, \rm{fm}^{-3}$ with a hyperbolic tangent stitching function (matching in $c_s^2(n_B)$). We then use the {\sc QLIMR} module~\cite{Conde-Ocazionez:2025agu, ReinkePelicer:2025vuh} to solve the TOV equations~\cite{Tolman:1939jz, Oppenheimer:1939ne} for a sequence of neutron stars. Configuration files for the entire workflow will be available upon publication.  Because we have neglected hyperons and quark degrees of freedom, these solutions represent a qualitative picture of neutron stars in this model.

Since our study has focused on properties that are most relevant around $\nsat$, the properties at larger densities have not been constrained by astrophysical observations, heavy-ion collision data, nor even by causality and stability.
Lorentz covariance of the Lagrangian by itself does not guarantee that the thermodynamic EoS obtained for a particular parameter choice is causal,as vector interactions are not only included in this work, but also broadly explored. Therefore, causality must be checked. 
In fact, some high-likelihood \cmf{} solutions do lead to  acausal EoSs at high density, while fulfilling vacuum and saturation satisfactorily. 
However, we do not exclude such solutions for two main reasons.  
First, adding hyperons or quark degrees of freedom may soften the high-density EoS and restore causality within the broader \cmf{} framework. 
Second, acausality in the nucleonic extrapolation may indicate that this truncation is no longer valid at such densities, for example because additional interactions become important or because the relevant degrees of freedom change, as in quarkyonic matter~\cite{Legred:2025aar, McLerran:2018hbz}.
We therefore stress that the neutron star solutions we show are in some sense ``extrapolations" of the nucleonic, mean-field EoS to high densities.

\begin{figure}
    \centering
    \includegraphics[width=0.99\linewidth]{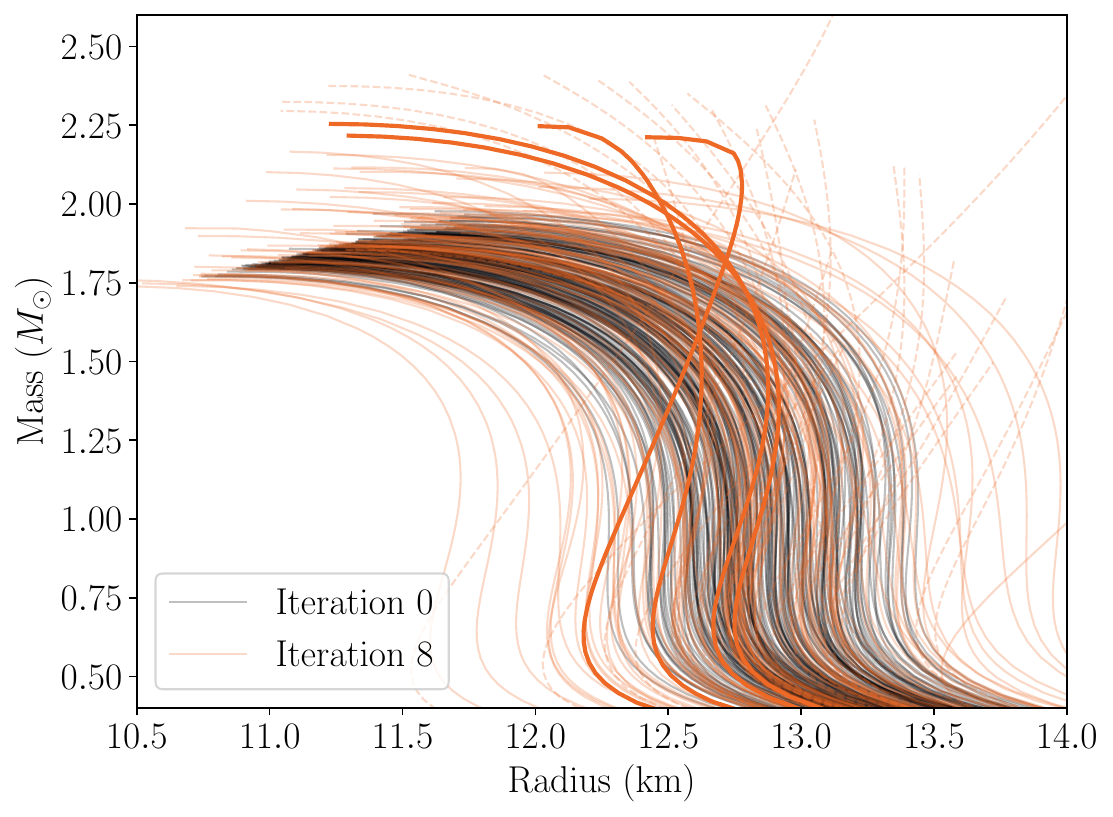}
    \includegraphics[width=0.99\linewidth]{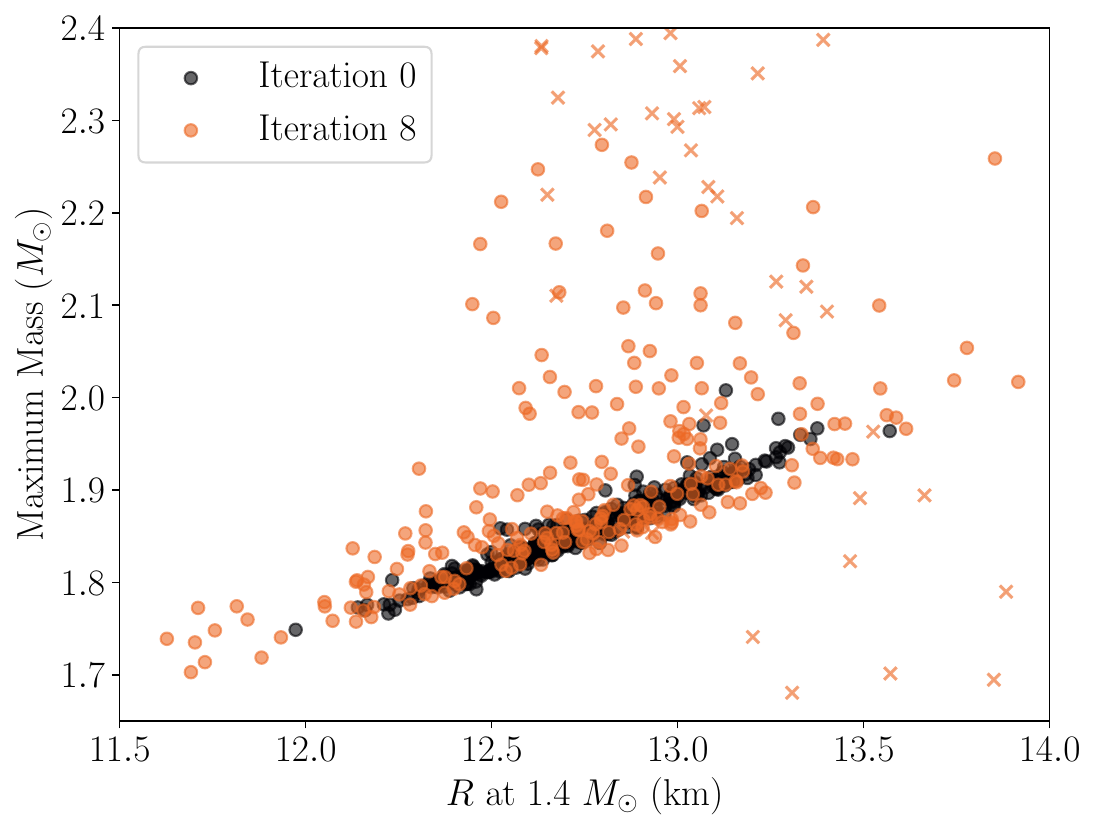}
    \caption{\emph{Top}:  mass-radius curves assuming the nuclear EoS holds over the entire NS (\emph{i.e.} assuming neutron star cores are purely nucleonic).  We display $\sim 200$ EoSs sampled from the posterior after iteration 0 (in black) and iteration 8 (in orange).  EoSs that become acausal below their TOV maximum mass (see text) are marked with dashed lines.  EoSs with $M_{\max}> 2.17\, M_{\odot}$ and $R_{1.4} < 13\,\mathrm{km}$ are highlighted with denser lines. \emph{Bottom:} the distribution of $M_{\max}$ vs $R_{1.4}$ for the posterior after iteration $0$ and $8$.    }
    \label{fig:neutron-stars}
\end{figure}

In \Cref{fig:neutron-stars}, we display properties of neutron stars for EoSs corresponding to posterior samples. In the top panel, we show randomly sampled mass-radius curves. We exclude EoSs which do not lead to thermodynamically stable solutions\footnote{In this case, thermodynamic instability means the baryon susceptibility $\chi_{BB} < 0$ or equivalently $dP/dn_B < 0$.} because such stars cannot be used to self-consistently solve the TOV equations. We do include EoSs that become acausal at a density below the maximum mass, but we mark them with dashed lines.    
We see that, even after iteration 0, there is already substantial spread in the range of neutron star masses and radii, and that after iteration 8 this general spread is still reflected.  In the bottom panel, however, we see that after iteration 8 there are weaker correlations between the maximum mass and the radius of a 1.4 solar mass neutron star ($M_{\max}-R_{1.4}$), under the posterior of iteration 8 versus the posterior of iteration 0.  We take this to indicate that as the parameter space is expanded, there will be a greater range of possible macroscopic neutron star behavior.  We also find this increased spread reflected in the number of acausal EoSs; at iteration 0, none of our EoSs was acausal, whereas at iteration 8  $\sim 15\% $ of solutions are acausal.

In particular, we draw three main conclusions from Fig.~\Cref{fig:neutron-stars}.  
First, when the \cmf{} parameter range is extended, more massive neutron stars are possible even with purely nucleonic EoSs.  
In the past massive neutron stars have been associated with a transition to quark matter (see e.g. \cite{Dexheimer:2020rlp}). 
Second, among EoSs which reach $2\,M_{\odot}$, their corresponding radii of  lower-mass $\sim 1 \,M_{\odot}$ stars are quite varied, over about $\sim 2\,\rm{km}$, which is important for measuring the EoS using  gravitational-wave and X-ray data.  
Third, some subset of these high likelihood EoSs that only contain nucleonic matter ``bend-back", which is due to the softening of the EoS at low densities, a phenomenon that has already been associated with particular vector self-interactions in the CMF model and others~\cite{Dexheimer:2018dhb}. This phenomena has also been associated with quarkyonic stars, or more generally deconfinement (see \emph{e.g.} \cite{McLerran:2018hbz,Tan:2021ahl}).  In general,  we find that the diversity of neutron stars which can be attained with nucleons alone will be enhanced by thorough exploration of parameter space, which means that ruling out nucleons in favor of other species depends on thorough parameter space exploration.  While acausal EoSs do indeed show some of the most extreme examples of each of these points, the phenomenon also appears for purely causal EoSs.

\subsection{Other choices of parametrizations of the quartic Lagrangian}
\label{sec:physical-couplings-constraints}

\begin{figure*}
    \centering
    \includegraphics[width=0.99\linewidth]{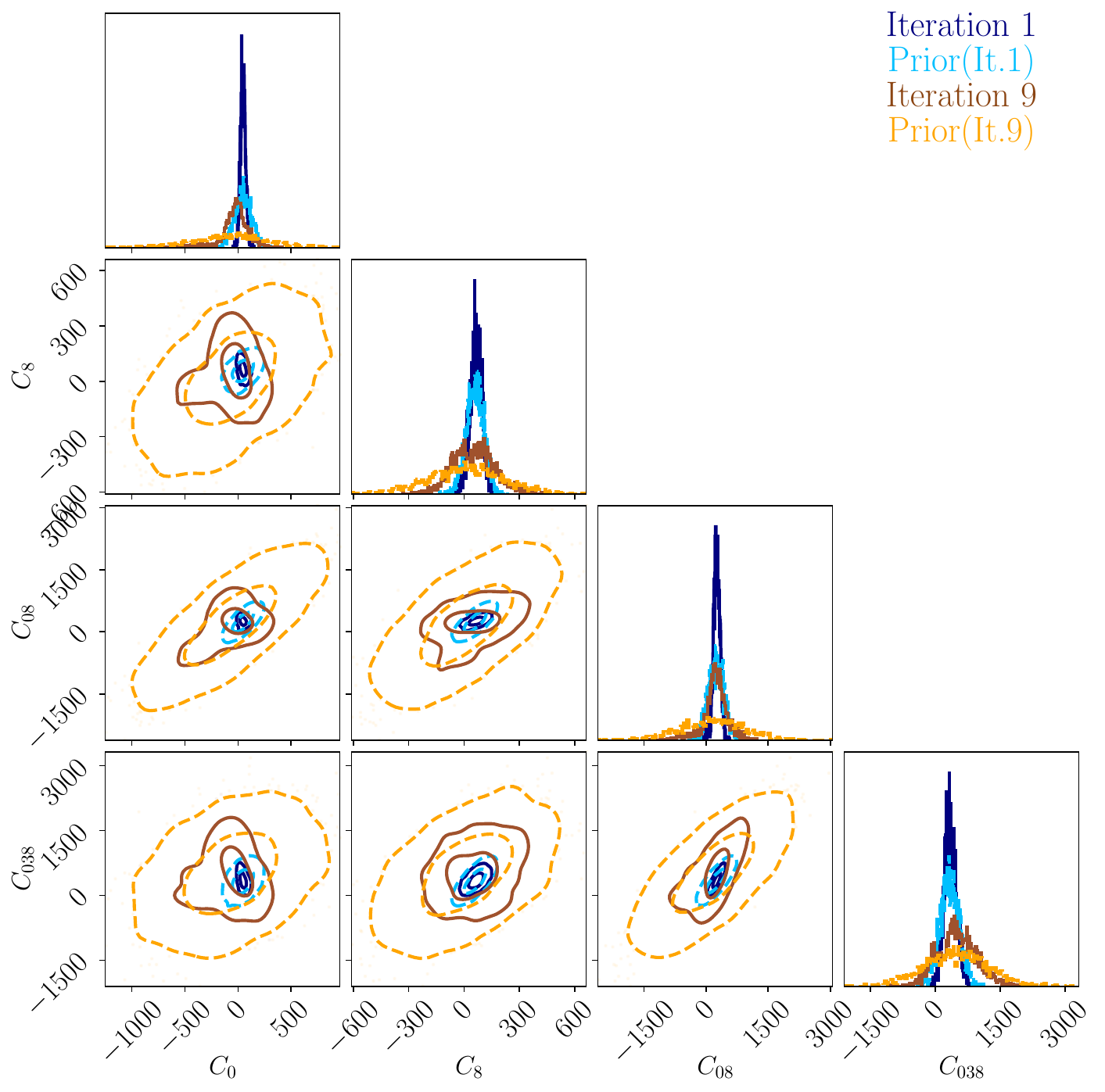}
    \caption{The induced posterior on the interaction coefficients written in the diagonal singlet-octet basis after iteration $1$ (blue) and $9$ in brown, with priors given by dashed lines in light blue and gold respectively.}
    \label{fig:physical-couplings-posterior}
\end{figure*}

We now discuss the induced constraints on the  couplings when writing the interactions in terms of the diagonal singlet-octet fields discussed in \Cref{sec:vector-couplings}. To do this, we merely transform our posterior and prior distributions using the mapping presented in \Cref{eq:physical-couplings-defs}.
We display a representative set of prior and posterior constraints in \Cref{fig:physical-couplings-posterior}.  We see that the pair of coefficients $C_{08}$--$C_{0}$ is particularly well-constrained relative to the prior.  However, all of the posterior 90\% regions still grow between iterations $1$ and $9$, indicating that more prior volume with valid solutions is being found.  Nonetheless, interpreting the relationships between the marginal priors and posteriors is subtle, since there may be higher dimensional correlations that are not captured in the $2$-dimensional posterior projections.  We believe that the diagonal singlet-octet basis couplings likely give a better description of the absolute constraints being imposed from constraints on nuclear matter, though our results are likely still, in some ways, prior-driven.   

\section{Discussion}
\label{sec:discussion}

In this work, we developed a neural-accelerated Bayesian framework for exploring the parameter space of chiral mean field models using vacuum baryon masses and nuclear saturation observables.  In the \cmfpp framework we used here, we explore 9 relevant free parameters that relate to these observables.  
We find that while at first glance the posteriors might not appear to be constraining, this can be misleading because the vast majority of parameter sets are incompatible with saturation properties. 
The constraint acts primarily on correlated combinations of \cmf{} parameters, not on most individual Lagrangian coefficients.  As a result, the viable region is sparse in the full parameter space, but broad when projected onto one or two parameters.

This distinction is very important when thinking about how the posterior should be used.  A marginalized corner plot is not a recipe for drawing new viable models. A random point inside the apparent range of the posterior is very likely to have a low likelihood.  The physically useful object is the set of posterior samples, together with the correlations among them.  For this reason, density-estimate-based visualizations, such as \Cref{fig:vector-self-coupling}, are useful diagnostics, but they do not contain all of the information needed for downstream calculations.  Posterior samples, sampler settings, and emulator configurations will therefore accompany this work~\footnote{To be released upon publication}, so that the inferred \cmf{} parameter sets can be used directly in future studies.

Taking our posteriors from our exploration and studying the realm of possible nucleon EoSs and how they affect astrophysical observables, we find a broad range of possible macroscopic behavior.  In particular, enlarging the explored parameter region weakens correlations, such as the one between $M_{\max}$ and $R_{1.4}$.  This is consistent with the general expectation that low-dimensional EoS parametrizations can induce correlations among macroscopic observables simply because they restrict the allowed functional space~\cite{Legred:2022pyp,Legred:2025aar}.  We emphasize, however, that the neutron-star calculation here is only an extrapolation of the nucleonic sector.  Hyperons and quarks are natural degrees of freedom in the \cmf{} model, and they can and will modify the high-density EoS.  This is why we do not impose astrophysical constraints in the present analysis, instead leaving that to future work.

The key difference between our approach here and previous Bayesian analyses in microphysical models is twofold. First, to our knowledge, this is the first Bayesian calibration performed using a full SU(3) \cmf{} model, in which the couplings are constrained by the underlying $\suthree{}$ structure, rather than a nucleonic RMF Lagrangian. Second, we introduce a generalized  vector self-interaction sector that respects the chiral symmetry of the model, significantly enlarging the parameter space.
Because of these extensions, the enlarged parameter space contains directions that are only weakly constrained by nuclear saturation observables, the data primarily constrain correlated combinations of the parameters, rather than individual Lagrangian coefficients. Consequently, the physically relevant information for \cmf{} is in the parameter correlation of the full posterior.
Opening up this phase space also led to new, never explored regimes that \emph{can} reach very heavy maximum masses at least for nucleonic only equations of state that was previously thought to be impossible within \cmf. Further exploration is still needed on the astrophysical properties of more exotic phases of matter in the core. 
Still, this study highlights the importance of not holding parameters fixed when conducting Bayesian parameter estimation because correlations between parameters can significantly affect conclusions.

Regarding the generic couplings, we allow a broader set of quartic vector self-interactions consistent with the chiral structure of the \cmf{} model.  In this basis, a single coupling does not usually correspond to a single physical channel.  Instead, changing one of the $g_4^{ij}$ coefficients changes several combinations of $\omega$, $\rho$, and strange-vector interactions at once.  Conversely, a physical channel, such as an $\omega^4$ or $\omega^2\rho^2$ interaction, depends on correlated combinations of the underlying chiral-invariant couplings. 
This difference is especially important when comparing to RMF parametrizations in which coefficients multiply physical channels directly.
This is why constraints that look weak in the original $g_4^{ij}$ basis can become more informative in the diagonal singlet-octet basis discussed in \Cref{sec:physical-couplings-constraints}.  
The data do constrain the model, but mainly through correlated combinations of the Lagrangian coefficients. Thus, the original coefficients $g_4^{ij}$ can have broad marginalized posteriors even when combinations such as the singlet--octet couplings are more tightly constrained.

This also explains why our results remain prior-dependent.  In a high-dimensional correlated model, the prior does not merely set harmless outer bounds, but rather it determines which correlated directions are allowed to participate in the inference.  The difference between our work and previous literature (\emph{e.g.}~\cite{Traversi:2020aaa,Malik:2023mnx,Scurto:2024ekq,Xie:2026hwg,Ma:2026dta}) is that, in the present work, the correlated directions are made explicit by the chiral-invariant parameterization, instead of being fixed by construction.  Thus, the broad marginal posteriors should not be interpreted as evidence that the framework is uninformative. We in fact show that the full posterior is likely highly structured, but in a way that reflects the complexity of the chiral interactions themselves.

A natural direction for future work is to refine each layer of the inference.
 The likelihood treats the saturation observables as independent Gaussian constraints, even though some of them, especially the symmetry energy and its slope, are known to be somewhat correlated.  The quoted uncertainties should therefore be interpreted as effective tolerances, not as a complete statistical model of nuclear data.  The vacuum baryon masses are also used as effective \cmf{} calibration targets chosen for consistency with previous \cmf{} studies, rather than as a fit to the most up-to-date PDG central values.  Finally, the in-medium inference relies on a neural-network emulator.  We have checked that the emulator is sufficiently accurate for the exploratory parameter-space mapping performed here, but a precision analysis should include direct recomputation or likelihood reweighting for the highest-likelihood samples.

This work raises several questions and possibilities for the future.
The next step is to add information that breaks the relevant degeneracies: correlated nuclear-matter constraints (see e.g. \cite{Drischler:2017wtt}), neutron-matter calculations, heavy-ion information (see e.g. \cite{Danielewicz:2002pu,Oliinychenko:2022uvy,Yao:2023yda}), and astrophysical observations, while consistently including the additional degrees of freedom that can appear at high density.  The framework developed here is designed specifically for that purpose.  Additional observables can be added to the likelihood, additional \cmf{} degrees of freedom can be included in the emulator, strangeness properties could be checked against recent chemical potential constraints \cite{Noronha-Hostler:2026ykk}, and posterior samples can be propagated into downstream neutron-star calculations without having to assume, in advance, which microscopic channel should be held fixed.

\section{Acknowledgments }

The authors are
supported by the NSF under the MUSES collaboration
OAC2103680.
V.D. acknowledges support from the U.S. Department of Energy, Office of Science, Nuclear Physics program under Grant DE-SC0024700 and from the National Science Foundation under grant  NP3M PHY2116686.
N.Y. acknowledges support from the Simons Foundation through Award No.~896696, the Simons Foundation International through Award No.~SFI-MPS-BH-00012593-01, and the NSF through Grant No.~PHY-25-12423.
J.N.H. acknowledges support from the US-DOE Nuclear
Science Grant No. DE-SC0023861.
We also acknowledge support from the Illinois Campus Cluster, a computing resource that is operated by the Illinois Campus Cluster Program (ICCP) in conjunction
with the National Center for Supercomputing Applications (NCSA), which is supported by
funds from the University of Illinois at Urbana-Champaign.
This work used the Delta system at the National Center for Supercomputing Applications through allocation PHY250117 from the Advanced Cyberinfrastructure Coordination Ecosystem: Services \& Support (ACCESS) program, which is supported by National Science Foundation grants $\#$2138259, $\#$2138286, $\#$2138307, $\#$2137603, and $\#$2138296.

\bibliography{inspire,NOTinspire}
\appendix
\crefalias{section}{appendix}
\crefalias{subsection}{appendix}

\section{CMF solutions with $m_0\neq 0$}
\label{app:mzero-nonzero}
We now discuss an additional inference where we do not assume the entire nucleon mass is generated dynamically. In practice, within the \cmf{} model this is controlled by a parameter $m_0$ that contributes a ``bare mass" correction to the hadron masses expressed in Eq.~\eqref{eq:baryon-masses}.  While such a correction does not explicitly break chiral symmetry in the non-linear realization, it quantifies how much of the baryon mass is not dynamically generated by the scalar condensates. While physically we expect such a term to be relatively small if most of the baryon mass is generated by the scalar condensates, we do not require it to be \emph{exactly} zero. Instead, we treat $m_0$ as a phenomenological parameter that can account for mass generation effects not captured by the condensates. Furthermore, systematic model error may indicate that a larger value of $m_0$ should potentially be tolerated.  In this appendix, we do not constrain $m_0$ to be small, but  in our relevant data product, $m_0$ is tabulated so that downstream analyses can make whatever cut is appropriate for their purposes. 

We find that introducing this parameter leaves the conclusions in our work qualitatively mostly unchanged, although it may lead to distinct phenomenology in neutron star mass-radius curves.  However, a technical challenge arises in sampling when $m_0$ is introduced as it is completely degenerate with the choice of $g_1^S$ in determining the baryon masses.  This can be seen by inserting Eqs.~\eqref{eq:nucleon-sigma-coupling} and \eqref{eq:nucleon-zeta-coupling} into~\eqref{eq:baryon-masses}, and recalling that in our framework $\zeta_0$ and $\sigma_0$ are assumed to be known exactly.  We briefly discuss how this technical challenge is handled in \Cref{app:m0-degenerate-sampling} before discussing results in \Cref{app:m0-degenerate-results}.

\subsection{Sampling scalar parameters when $m_0\neq0$}
\label{app:m0-degenerate-sampling}
\begin{figure*}
    \centering
    \includegraphics[width=0.9\linewidth]{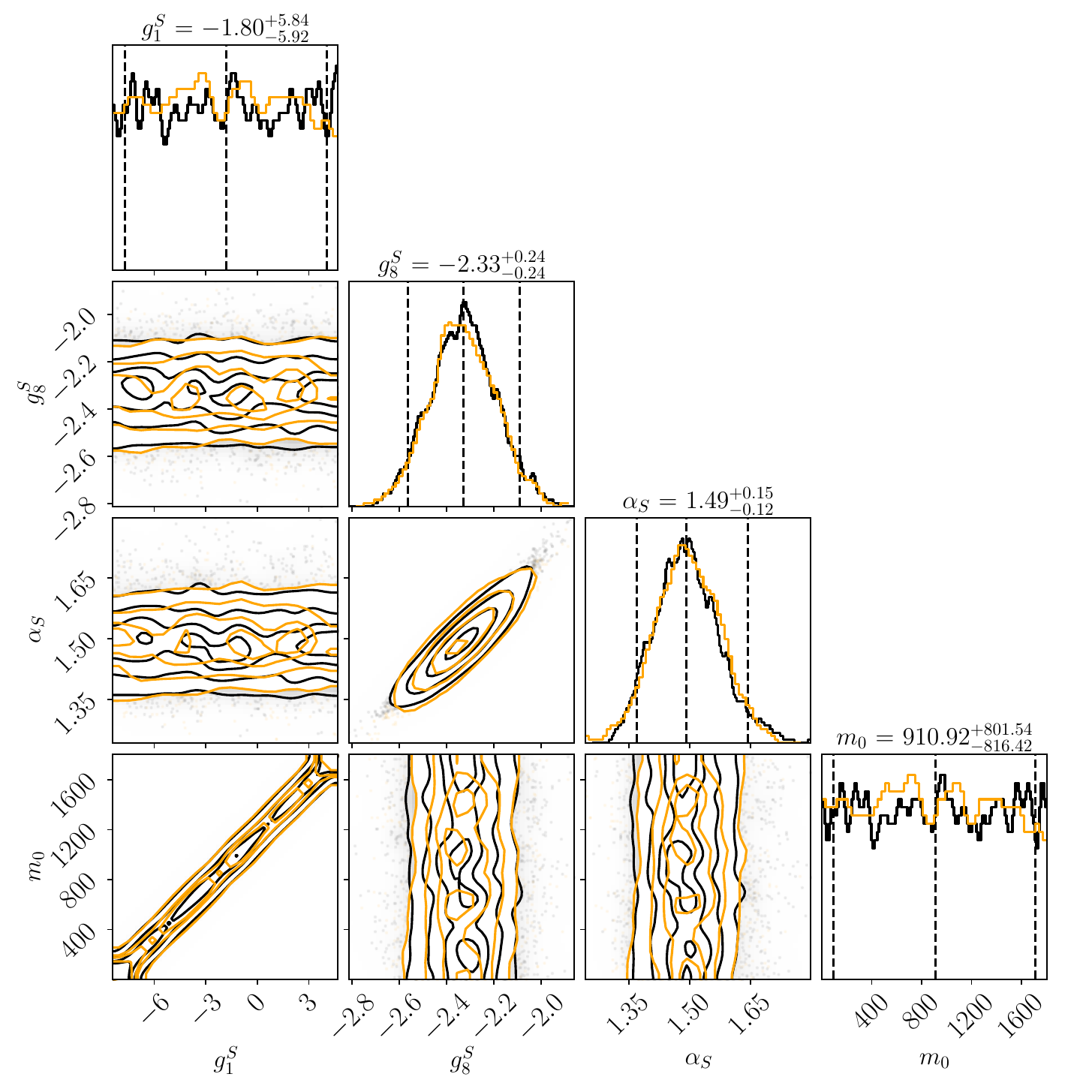}
    \caption{Same as \Cref{fig:vacuum-parameters-corner}, but allowing $m_0$ to vary.  The pure degeneracy with $g_1^S$ is seen in the bottom left corner.  By performing a change of variables (see the text), we can effectively model the entire distribution (in gray) by the product of a multivariate Gaussian distribution and a uniform distribution (in yellow).}
    \label{fig:scalar-nucleon-any-m0}
\end{figure*}

Because $m_0$ and $g_1^S$ enter into all expressions for masses linearly, they are completely degenerate with a constant of proportionality that depends only on the  values of $\zeta_0$ and $\sigma_0$. This is not, by itself, a serious problem, as we find the nested sampler is capable of identifying the full solution set.  We display the inferred posterior distribution in \Cref{fig:scalar-nucleon-any-m0} in gold. As can be seen, the parameter $g_1^S$ is no longer effectively modeled as a component of a multivariate Gaussian along with $g^S_8$ and $\alpha_S$, as the correlation with $m_0$ has smeared out the relation.  However, the variable $ y^S\equiv g_1^S - m_0 /\kappa$ is still approximately normally distributed for $\kappa =  -\left(\sqrt{2/3}\sigma_0 + \sqrt{1/3}\zeta_0\right) \approx 139$ (determined again by the choice of $\zeta_0$ and $\sigma_0$, though for any value near this value the distribution will remain mostly Gaussian).   We fit a Gaussian to the joint distribution $y^S, g^S_8, \alpha_S$, and treat the $m_0$ distribution as uniform for use in the inference.  In \Cref{fig:scalar-nucleon-any-m0}, we map the Gaussian distribution back to  $g^S_1, g^S_8, \alpha_S,$ and $m_0$ (in gray) to show that the fit (in yellow) is acceptable.

We proceed as described in \Cref{sec:bayesian_analysis}, but allowing for $m_0$ to be a free parameter in the computation of the in-medium matter properties.  It is therefore also treated as a free parameter by the emulator. We repeat the analysis performed in the main body of the text with the additional free parameter $m_0$.

\subsection{Results when $m_0\neq 0$}
\label{app:m0-degenerate-results}

\begin{figure*}
    \centering
    \includegraphics[width=0.99\linewidth]{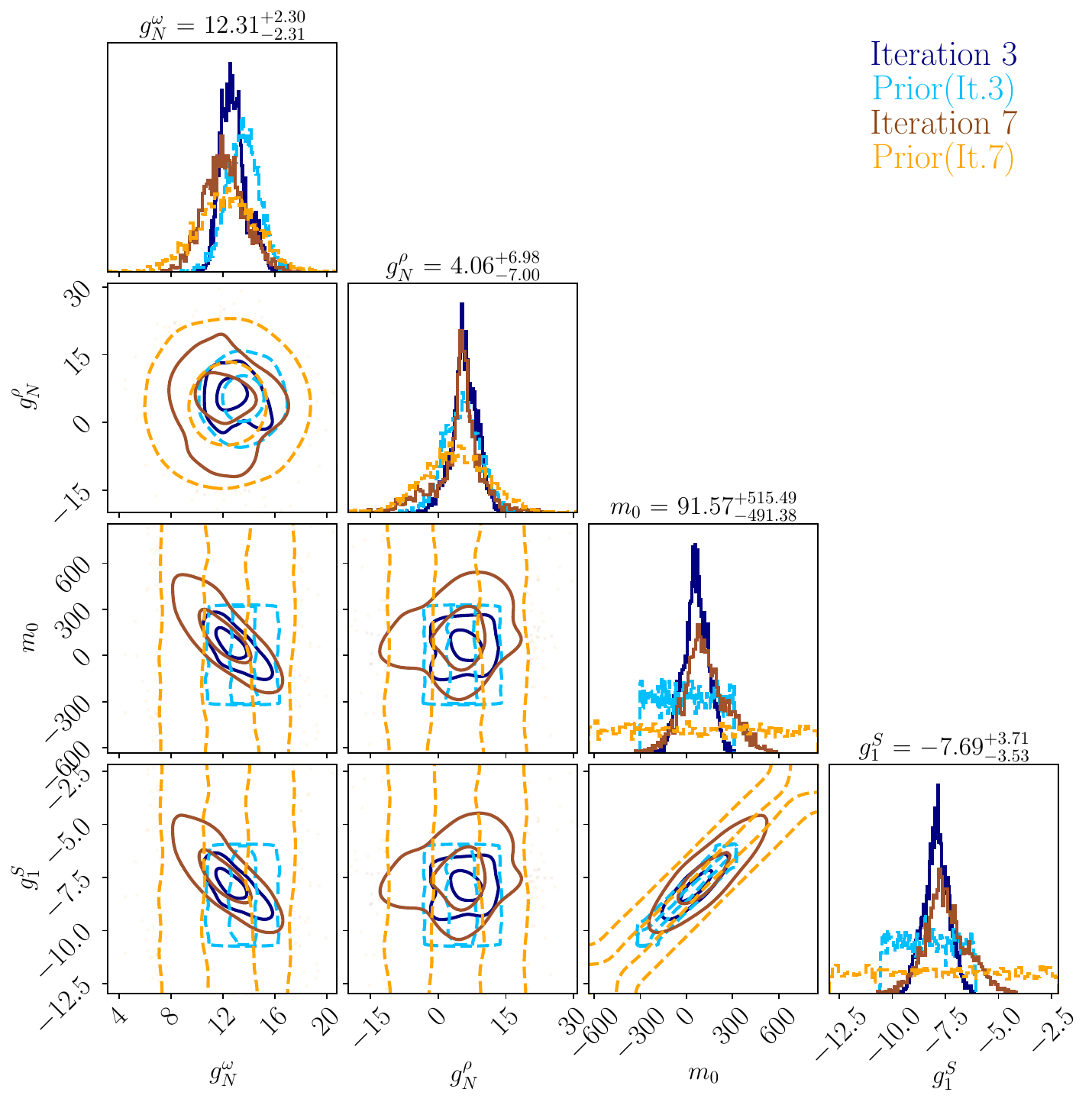}
    \caption{Posterior distribution after iteration 3 (blue) and iteration 8 (brown) on the \cmf{} parameters, which we find are most affected by the inclusion of $m_0$ as a free parameter.  Priors are marked in dashed lines of light blue and gold respectively}
    \label{fig:corner-any-m0}
\end{figure*}

We find that the results of the full inference using the emulator are, for the most part, qualitatively unchanged when looking just at the inferred large-scale structure in the \cmf{} parameter posteriors.  However, as noted, the large-scale variation in the \cmf{} posterior is not reflective of the true variation in the likelihood; it may be the case that at small scales the results are substantially different, though this is difficult to quantify. We display the posterior after iterations 3 and 8 in \Cref{fig:corner-any-m0}.  Our inference gradually broadens the parameter on $m_0$ (which due to correlations broadens the posterior on $g_1^S$).  The parameters $m_0$ and $g_1^S$ are correlated with $g_{N\omega}$ via the likelihood on in-medium observables, indicating that the inclusion of $m_0$ has a nontrivial effect on what combinations of \cmf{} parameters can reproduce saturation properties of matter.  

\begin{figure}
    \centering
    \includegraphics[width=0.99\linewidth]{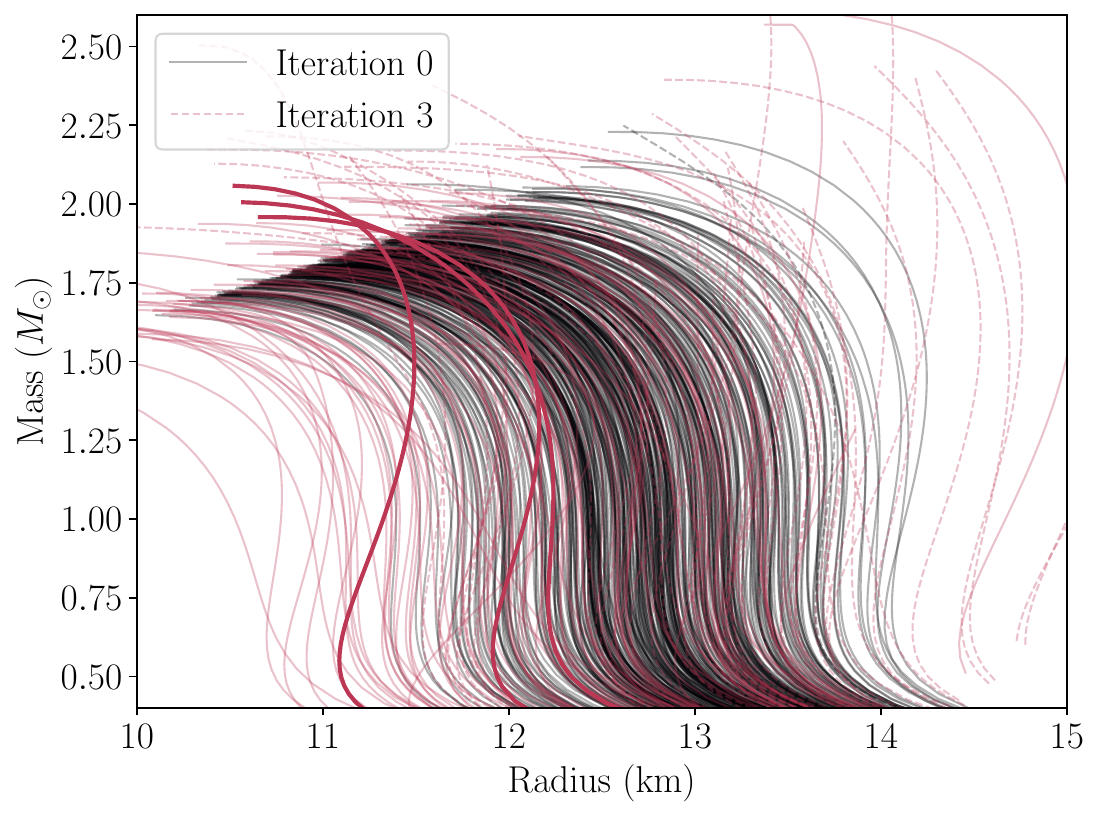}
    \includegraphics[width=0.99\linewidth]{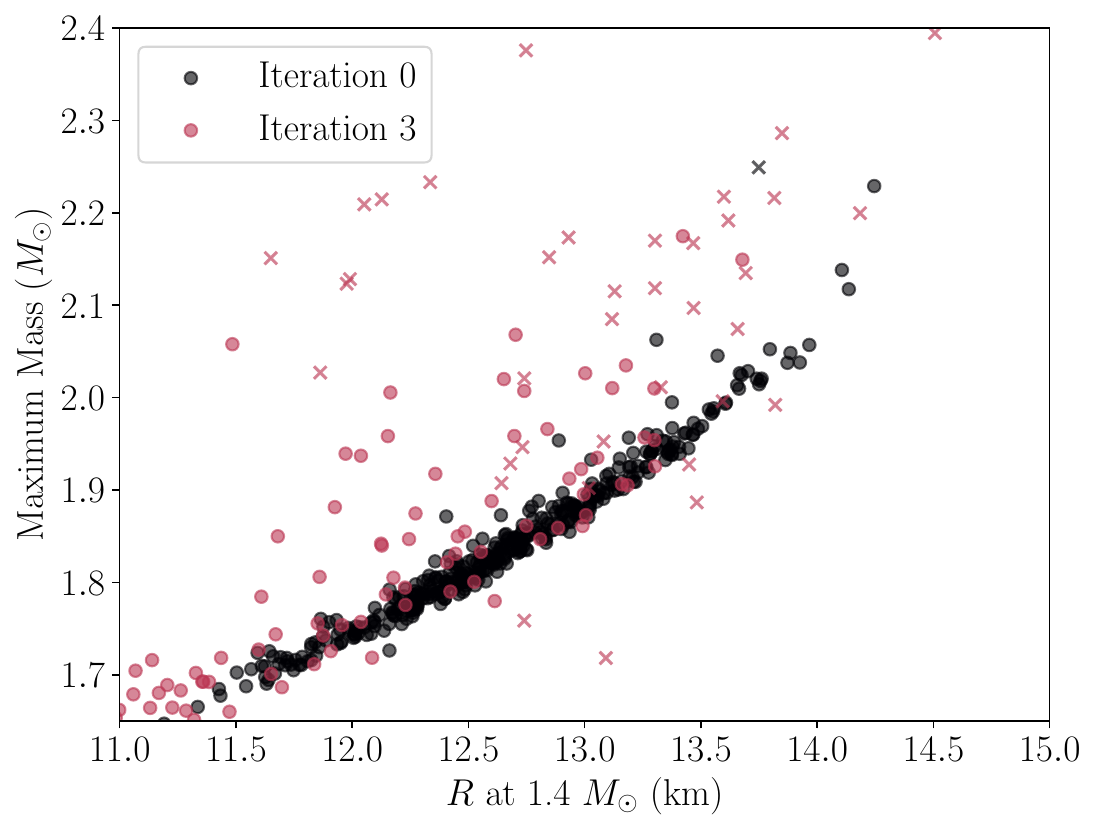}
    \caption{Analogous to \Cref{fig:neutron-stars} but for the runs where $m_0$ is allowed to vary away from zero.  We show representative $M-R$ curves from iteration $0$ in black and from iteration $3$ in maroon.  EoSs with $M_{\max} > 1.95\,M_{\odot}$, but $R_{1.4} < 12.5\, \mathrm{km}$ are highlighted with denser lines.  }
    \label{fig:neutron-stars-nonzero-m0}
\end{figure}

Extrapolating these differences allows us to examine the mass-radius curves of neutron stars with EoSs with nonzero $m_0$.  We show the corresponding mass-radius curves, and the correlation between $M_{\max}$ and $R_{1.4}$ in \Cref{fig:neutron-stars-nonzero-m0}.  Even after iteration $3$, we now see substantial differences relative to iteration $0$.  Most importantly, we see larger values of $M_{\max}$ for given values of $R_{1.4}$.  In \Cref{fig:neutron-stars-nonzero-m0}, we highlight mass-radius curves with $M_{\max} > 1.95\,M_{\odot}$, which also have small radii $\sim 11$---$12\,\mathrm{km}$.   However, the introduction of $m_0\neq 0$, while producing some promising neutron stars, also leads to many highly unphysical EoS candidates that are nonetheless consistent with constraints on nuclear matter near saturation.  Therefore, at higher iteration numbers, it is increasingly challenging to find a large sample of physical EoSs, though a handful of candidates are potentially interesting from an astrophysical point of view. However, we stress again that this analysis represents an extrapolation of the nucleon EoS to the densities of neutron star cores.   This may not be valid if hyperons appear, or there is a phase transition to quark degrees of freedom.  Any robust astrophysical analysis should take into consideration these possibilities.  

\section{Decomposition in terms of \physcoupling{} parameters}
\label{app:phys_couplings}
In this appendix, we write the vector self interaction terms in \emph{\physcoupling{}} couplings;  coefficients correspond to the quartic terms in the Lagrangian multiplying particular interactions of the physical vector mesons such as $\omega^4$ and $\omega^2 \rho^2$.  We now give the expressions for these \physcoupling{} couplings in terms of the \Ccoupling{} couplings, which are constrained above.  Note that there are many more \physcoupling{} terms than nonzero \Ccoupling{} terms, because the \physcoupling{} terms mix the singlet and octet states, leading to a larger set of possible interaction terms, of which the couplings are related by chiral symmetry by the expressions below:

\begin{align}
g_{\omega^4}^{\rm eff}
&=
\frac{4}{9}\mathcal{C}_0
+\frac{1}{9}\mathcal{C}_8
+\frac{2}{9}\mathcal{C}_{08}
-\frac{\sqrt{2}}{27}\mathcal{C}_{038},
\\
g_{\phi^4}^{\rm eff}
&=
\frac{1}{9}\mathcal{C}_0
+\frac{4}{9}\mathcal{C}_8
+\frac{2}{9}\mathcal{C}_{08}
+\frac{2\sqrt{2}}{27}\mathcal{C}_{038},
\\
g_{\rho^4}^{\rm eff}
&=
\mathcal{C}_8,
\\
g_{\omega^2\rho^2}^{\rm eff}
&=
\frac{2}{3}\mathcal{C}_8
+\frac{2}{3}\mathcal{C}_{08}
+\frac{\sqrt{2}}{3}\mathcal{C}_{038},
\\
g_{\phi^2\rho^2}^{\rm eff}
&=
\frac{4}{3}\mathcal{C}_8
+\frac{1}{3}\mathcal{C}_{08}
-\frac{\sqrt{2}}{3}\mathcal{C}_{038},
\end{align}
\begin{align}
g_{\omega\phi\rho^2}^{\rm eff}
&=
-\frac{4\sqrt{2}}{3}\mathcal{C}_8
+\frac{2\sqrt{2}}{3}\mathcal{C}_{08}
-\frac{1}{3}\mathcal{C}_{038},
\\
g_{\omega^3\phi}^{\rm eff}
&=
\frac{8\sqrt{2}}{9}\mathcal{C}_0
-\frac{4\sqrt{2}}{9}\mathcal{C}_8
-\frac{2\sqrt{2}}{9}\mathcal{C}_{08}
+\frac{5}{27}\mathcal{C}_{038},
\\
g_{\omega\phi^3}^{\rm eff}
&=
\frac{4\sqrt{2}}{9}\mathcal{C}_0
-\frac{8\sqrt{2}}{9}\mathcal{C}_8
+\frac{2\sqrt{2}}{9}\mathcal{C}_{08}
-\frac{2}{27}\mathcal{C}_{038},
\\
g_{\omega^2\phi^2}^{\rm eff}
&=
\frac{4}{3}\mathcal{C}_0
+\frac{4}{3}\mathcal{C}_8
-\frac{1}{3}\mathcal{C}_{08}
-\frac{\sqrt{2}}{9}\mathcal{C}_{038}.
\end{align}

\section{Additional tests of the saturation properties emulator}
\label{app:stress-tests}

We now discuss additional diagnostic tests of our emulator of saturation properties.  In general, we evaluate the emulator to be effective if emulator errors are typically close to or smaller than the uncertainties in the saturation properties.  This is acceptable here because we focus on parameter \emph{exploration}.  If a careful quantification of the parameter space, especially with respect to outliers, was required, likely stricter requirements would need to be set. For example, a natural criterion to use for robust inference would be that the median error in the emulator be much smaller than the uncertainty in the given variable. 

One simple diagnostic test we perform is that the emulator performs better in regions of parameter space where it has access to more training data. In the case of the emulator in the main text, the initial set of training data is clustered around the C1 configuration of \cmf{} parameters, and expanded relative to this point.  Therefore, the regions of parameter space closest to the C1 point should have the most training data, and we expect the emulator to be the most precise in this region.  In order to quantify this, we consider the Euclidean distance to the C1 point to be 
\begin{equation}
    D(x, \text{C1}) = 
    \left[{\sum_{I} \quant{\frac{g_4^{I, x} - g_4^{I, \text{C1}}}{\sigma_{g_4^{I
    , x}}}}^2}\right]^{1/2}
\end{equation}
where I ranges over the different indices of couplings in the Lagrangian of Eq.~\eqref{eq:couplings-definitions} (e.g. $I=220, 04, 22, 13 $), and $\sigma_{g_4}^{I,x}$ is the measured standard deviation of the parameter in the particular set of points that is being evaluated.

\begin{figure}
    \centering
    \includegraphics[width=0.99\linewidth]{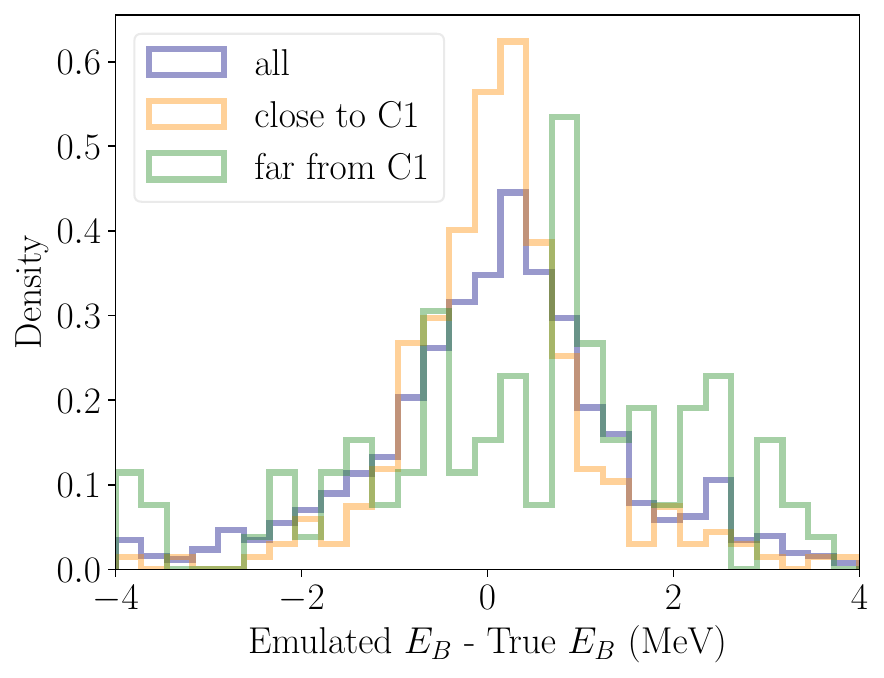}
    \caption{The difference between the predicted and true binding energy of matter at saturation density for a large set of unseen training samples (see text) for points that are ``close" (orange) and ``far" from the initialization (C1) point in parameter space (green, see text). The overall distribution is overplotted in blue.}
    \label{fig:be-vs-distance-to-C1}
\end{figure}

In \Cref{fig:be-vs-distance-to-C1} we show the error in the emulator prediction of the binding energy of saturated matter for all of the training data that was generated in iterations 4-11, with the emulator that was used in iteration 4.  That is, none of the points being tested have been seen by the emulator as training data, and in fact in much of the parameter space the emulator has very sparse training data, consisting of only a handful of points.  This figure shows the full set of emulator errors (blue), the errors in that have $D(x, \text{C1})<2$ (orange), those that have $D(x, \text{C1}) > 10$ (green).  That is, the orange points are ``close" to where the iterative process was initialized, while the green points are ``far" and the emulator should in principle have less training data.  Surprisingly, in both cases most of the emulator errors are small, in that the difference between the predicted and true values is $\lesssim 1.8\, \rm{MeV}$ (which is $3$--$\sigma$ compared to our estimate of the current measurement uncertainty). However, for points far from the C1 point, there is a much higher density of points with very large errors ($\gtrsim 2\,\rm{MeV}$) (\emph{c.f.} green to orange in the tails).  Therefore, we see that as expected, points near the starting point of the iteration scheme, where the emulator has access to abundant training data, are better predicted by the emulator.  However, the emulator is still surprisingly good when extrapolated outside its training data. 

\section{Toy model of a qualitatively similar system}
\label{app:toy-model}
We here introduce a lower-dimensional and simple toy problem to explain the structure of the likelihood we found in this paper.  Consider a mapping 
\begin{align}
    y &= f(x_1, x_2),\\
    f(x_1, x_2) &= \sin(x_1 + x_2),
\end{align}
with a simulated measurement $d$ giving
$y_1 = 1.0 \pm 0.2$ (where uncertainties are assumed Gaussian).
The inference problem for
$P(x_1, x_2|d)$ has a likelihood that is highly structured in $\{x_1, x_2\}$, though as we show below, the induced marginal posteriors on $x_1$ and $x_2$ are effectively Gaussians. 

\begin{figure}
    \centering
    \includegraphics[width=0.99\linewidth]{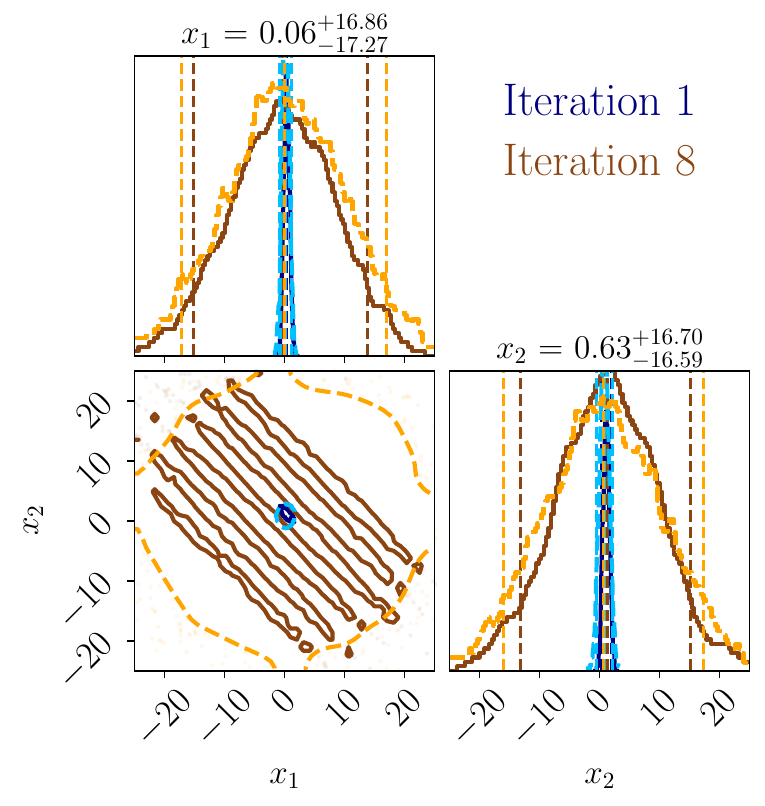}
    \caption{The toy model posterior described in the text after iteration 1 (blue) and 8 (brown). Priors are plotted in dashed lines of lighter shades.}
    \label{fig:corner-toy-model}
\end{figure}

Points will achieve high likelihood if $x_1 + x_2 \approx 2n\pi + \pi/2$ for integer $n$, so the likelihood does not have compact support, and it is not possible to choose a prior that completely encloses the region of nontrivial likelihood. We use the same iterative inference scheme as introduced in \Cref{sec:bayesian_analysis} to initialize a small prior range and expand to broader priors. We plot the posterior distribution after iteration $1$ (blue) and iteration $8$ (brown) in \Cref{fig:corner-toy-model}.  We see that despite the fact that the 2-D posterior is highly structured, the 1-D marginal posteriors are nearly gaussian, and indeed look almost indistinguishable from the priors (dashed lines in lighter shades).  In the case of the main text, we expect that the highly structured likelihood is only visible in dimension $\geq 5$, and any lower-dimensional projection will be dominated mostly by the prior.   In contrast to the case of the likelihood in the main text, we expect this case to show nontrivial homology groups (in particular, nontrivial $H_0$).  We verify this using our persistent homology code.

\section{Tests of Structure in the Likelihood}
\label{app:data-structure}
We here give more detailed descriptions of the tests implemented in the main text for attempting to identify structure in the full posterior distribution.  In general, our tests fall into two classes, which are in fact closely related. First are tests of clustering: we seek to identify if there is any structure to how points are distributed, which would point to ``islands" in the higher dimensional parameter space.  Second are tests of ``voids" in the data, which search for regions that \emph{do not} contain solutions, in such a way that these regions form ``holes" in the posterior.  We find null results for both classes of tests, indicating that our posterior distributions are consistent with being topologically trivial when probed at long length scales (that is, length scales comparable to the width of the posterior). 

\begin{figure}
    \centering
    \includegraphics[width=0.99\linewidth]{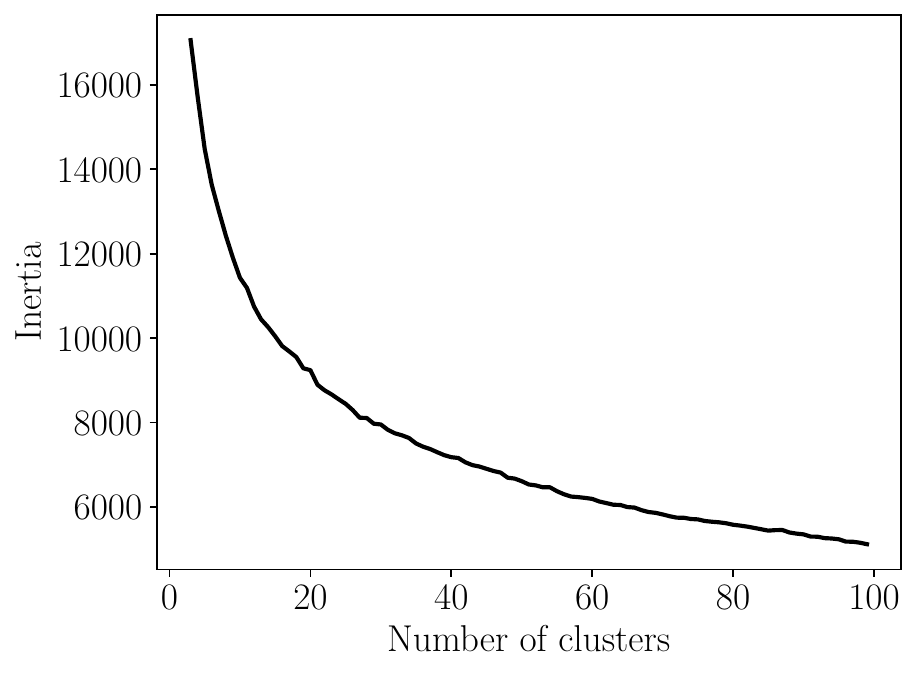}
    \caption{Inertia of a k-means clustering fit as a function of the number of clusters.  We use a ``typical iteration" of our algorithm, iteration 8 for this example.}
    \label{fig:inertia-vs-clusters}
\end{figure}

For tests of clustering, we first apply a k-means clustering algorithm to our full posterior, which is whitened by subtracting off the posterior mean and dividing out the posterior standard deviation for each variable.  Using  {\sc scikit-learn} implementations, we then compute a silhouette score for the clustering predictions.  We use $k\in [3,100]$ clusters, and compare silhouette scores, generating a so-called ``elbow plot" for the inertia, or ``within-cluster-sum-of-squares" of the fits, which we display in \Cref{fig:inertia-vs-clusters}.  In general, a smaller inertia indicates each cluster represents a more coherent class of examples, and therefore the clustering algorithm produced a ``better fit".  However, by definition, fits with more clusters will have less within-cluster variation.  If there is a clear point where the clustering algorithm seems to saturate in inertia as the number of clusters is increased, it indicates that the appropriate number of clusters has been found  and an ``elbow" forms in the plot.  We see no obvious elbow; it appears as if predictive success is still increasing with number of clusters even at $k=100$ clusters. The silhouette score, which measures how well-separated clusters are, is found to be effectively independent of number of clusters, at a nearly constant value of $\sim 0.12$, indicating that clusters are not well separated.

Another measure of clustering is provided by our second test of data structure, which is computing the persistent homology of the Rips-Vietoris complex formed from  posterior point-cloud data\footnote{This approach frequently goes by the name topological data analysis, or TDA}.   Technically, persistent homology is the simplicial homology of a family of simplicial complexes that form a filtration.  In geometric terms, a filtration is a sequence of nested topological spaces that (in this case) allow efficient computation of homology groups.  Homology groups are defined by defining so-called ``boundary" operators on topological spaces, and measuring the extent to which, at a certain dimension $d$, all $d$-dimensional ``cycles" within the space coincide with boundaries of $d+1$ dimensional structures.  If there are cycles that are not the boundary of a higher dimensional structure, that cycle is called a ``generator" of the dth homology group, $H_d$.  As an example, a circle contains a single distinct cycle, which is the circle itself, which is not the boundary of a higher dimensional structure.  Therefore the circle has one ``hole" at dimension 1.  The unit disk on the other hand, has the same cycle, but in this case the cycle corresponds to the boundary of the disk, therefore a disk has no topological ``holes".  

The persistent homology of a dataset is computed by first building topological spaces that represent the dataset. This is done by using a \emph{filtration}, or sequence of nested complexes of topological spaces.  In this case, the complex is a simplicial complex, which means it is a topological space composed of simplices that are generalized triangles in higher (or lower) dimensions.  A complex is computed by considering sets of points, and inserting a simplex between them if they are sufficiently closely grouped, based on some threshold distance.  For example, two points will be connected by a line if they are less than a distance $x_{\rm thresh}$ apart.  The crucial point is that as the closeness threshold is relaxed, more and more points will be ``connected" by various topological structures, but each topological space will naturally contain, as a subspace, every space that was computed at a stronger threshold (since there were strictly fewer components).  Thus, at each distance, we get a topological space of which we can compute the homology, and holes that appear at some distance and \emph{persist} over a large range of distance thresholds, are the most likely to be physically meaningful.
Therefore, generically,   persistent homology is a strategy for computing the number of ``holes" in a dataset. 

However, for dimension $0$,  homology actually computes the number of distinct connected components of the topological space.  Therefore, the ``zero-dimensional" homology group ($H_0$) is also a measure of clustering.  If there were a handful of well separated clusters, these would be disconnected so long as the distance separation threshold $(x_{\rm thresh})$ is less than the distance separation of the clusters. 

\begin{figure}
    \centering
    \includegraphics[width=0.99\linewidth]{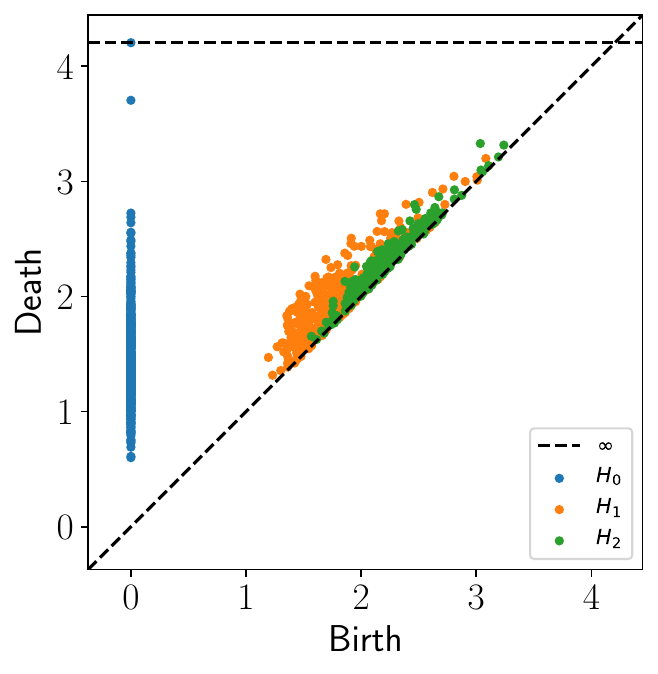}
    \includegraphics[width=0.99\linewidth]{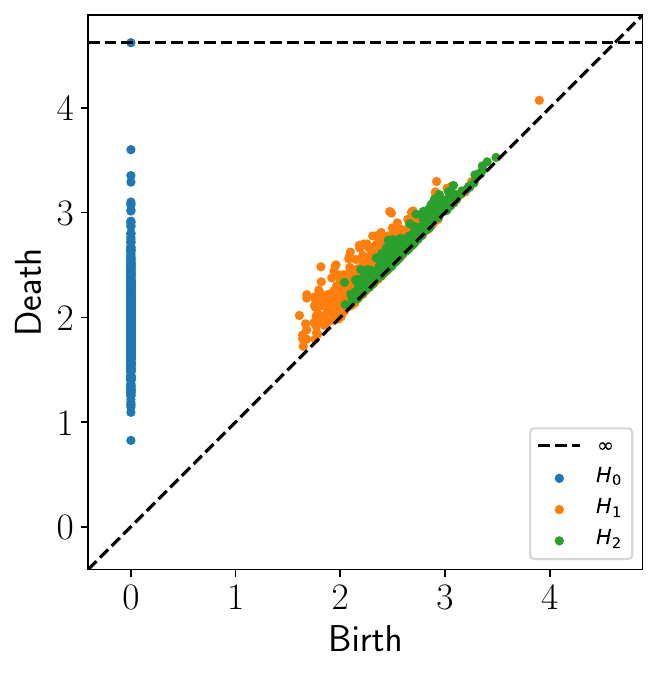}
    
    \caption{Persistence barcodes of the posterior of the 11th iteration of the algorithm presented in the main text, top, vs the same for data drawn from a multidimensional Gaussian distribution.
    }
    \label{fig:tda-barcodes}
\end{figure}

We use {\sc ripser} implemented with python bindings to compute the persistent homology of the posterior of the 11th iteration of our algorithm. We display the so-called ``persistence barcode" in \Cref{fig:tda-barcodes}.  This diagram shows the threshold length scale at which particular generators of homology first appear (birth) and last appear (death) as the distance threshold is raised.  We include in the same figure an example of the same barcodes generated using point-cloud data generated from a multivariate Gaussian distribution. We can see no distinction between the two barcodes by eye, which leads us to conclude that topologically our dataset is effectively identical to a Gaussian distribution in the sense that it has no meaningful 1-D loops or 2-D voids, and no meaningful clusters of points (besides the dominant cluster that is the distribution itself).   

We conclude that, at the sampled resolution and with the whitened Euclidean metric used here, the posterior shows no evidence for well-separated large-scale clusters or persistent topological holes. This does not exclude correlated degeneracy structure or likelihood variation on scales below those probed by the TDA calculation.  It is possible structure exists in higher dimension, though quantifying this is more computationally expensive.  Even if there is higher dimensional structure, it may be modified by the inclusion of additional degrees of freedom.  The emulator approach does not require understanding the high-dimensional structure \emph{a priori}, which is one of the benefits of using it to explore the parameter space.

\section{Algorithm of the saturation properties module}
\label{app:sat_module}

The {\sc Saturation Properties} module receives a tabulated nuclear equation of state  on a discrete grid of baryon density $n_B$ and charge fraction $Y_Q = n_Q/n_B$. The grid is not required to be regular in either variable, but it must be in MUSES format~\cite{ReinkePelicer:2025vuh,Jahan:2026hvs}. The algorithm first computes the saturation density, $\nsat$, using a Levenberg--Marquardt minimization which works as a root-finder to determine where $P(\nsat, Y_Q=0.5) = 0$. To avoid near-vacuum solutions where the pressure may satisfy $P\simeq 0$, rows below a user-configurable minimum density $n_B^{\min}$ are discarded. The default value is $n_B^{\min} = 0.05$ fm$^{-3}$. 

The nucleon mass must be supplied in the configuration file for the module to report the binding energy $\Esat$.  The other saturation properties, defined in can be computed either by fitting the energy expansion to the tabulated EoS or by taking numerical derivatives. The user can choose between the available methods.

In the \emph{fitting} method, beyond the five quantities used as likelihood constraints in this work ($\nsat$, $\Esat$, $K_0$, $\Esym$, $\Lsym$), the module also determines the next two orders of the isoscalar and isovector density expansions, $Q_0$, $Z_0$, $K_{\rm sym}$, $Q_{\rm sym}$, and $Z_{\rm sym}$, which we report here as additional characterizations of the EoS. Introducing
\begin{equation}
x=\frac{n_B-\nsat}{3\nsat},
\end{equation}
the density expansions implemented by the module, extending \Cref{eq:snm-expansion,eq:symmetry-energy-expansion}, are
\begin{equation}
E_{\SNM}(n_B) = \Esat
+\frac{1}{2}K_0x^2
+\frac{1}{6}Q_0x^3
+\frac{1}{24}Z_0x^4,
\end{equation}
and
\begin{equation}
S(n_B) = \Esym+\Lsym x
+\frac{1}{2}K_{\rm sym}x^2
+\frac{1}{6}Q_{\rm sym}x^3
+\frac{1}{24}Z_{\rm sym}x^4.
\end{equation}
Here, $K_0$, $Q_0$, and $Z_0$ are respectively the second-, third-, and fourth-order density coefficients of symmetric nuclear matter, while $\Esym$, $\Lsym$, $K_{\rm sym}$, $Q_{\rm sym}$, and $Z_{\rm sym}$ are the corresponding coefficients of the symmetry energy. 
To compute them, the module performs a nonlinear least-squares fit of the energy-per-baryon expansion to all data points within a configurable window around $\nsat$ and $Y_Q=0.5$. The fit is performed with the Ceres solver~\cite{Agarwal_Ceres_Solver_2022} using automatic differentiation, and the bulk parameters are extracted simultaneously from a single global fit.
In the \emph{derivative} method, the isoscalar properties are obtained from numerical derivatives of the energy per baryon along $Y_Q=0.5$. The module can use either a GSL natural cubic spline, which provides the function itself and second derivative, $\Esat$ and $K_0$, or central finite-difference stencils. The symmetry energy $\Esym$ and its slope $\Lsym$ are computed from derivatives of the energy per baryon in the  $Y_Q$ direction at $Y_Q=0.5$, using a fourth-order central finite-difference stencil when the required grid points are available. If the EoS table does not extend beyond $Y_Q=0.5$, the module automatically switches to a second-order backward stencil, with the step size chosen from the $Y_Q$ grid so that evaluation points coincide with existing grid points. The derivative method does not provide the third and fourth-order coefficients of \Cref{eq:energy-expansion} in the current implementation, these quantities are reported as \texttt{NaN} when this method is selected.

The module has been tested on existing \cmf{} and RMF parametrizations, and in realizations of a metamodel EoS~\cite{Margueron:2017eqc,Margueron:2017lup}, in which the input are the saturation properties themselves, providing a clean validation of the module. For tables regularly gridded in the $(n_B, Y_Q)$, both the derivative and the fitting methods converge to within 1\% error as long as the parabolic behavior around saturation density is captured.  

For tables regularly gridded in $(\mu_B,\mu_Q)$, the fitting method has a similar accuracy to grids regular in $(n_B, Y_Q)$, as long as the parabolic behavior is captured in the data. This typically requires step sizes of $\Delta \mu_B \lesssim 2$~MeV and $\Delta \mu_Q \lesssim 2$~MeV in our tests. The case of finite differencing is subtle. Near the liquid-gas transition, the map from
$(\mu_B,\mu_Q)$ to $(n_B,Y_Q)$ can become highly nonuniform because its Jacobian becomes very large. Since the derivatives are evaluated in $(n_B, Y_Q)$ using a 2-dimensional interpolation built with a Delaunay triangulation, if the step size is too small, all stencil evaluations fall on the same triangle, and any information about curvature of the EoS is lost.
If the step is too large, the stencil no longer probes the local curvature around saturation. Since the optimal step is grid dependent and no single value works across different $(\mu_B,\mu_Q)$ spacings, the fitting method is recommended for tables gridded in chemical potentials. In this work, we have used $\Delta \mu_B = 0.1$ and $\Delta \mu_Q = 1.0$~MeV.

This module can be used through the MUSES Calculation Engine~\cite{calculation_engine_all_versions}. The released version is archived on Zenodo~\cite{pelicer_2026_21079718} and the source code is available under the GPL-3.0 license in the public Git repository~\cite{muses_saturation_properties_repo}.

\end{document}

%% file: macros.tex
\newcommand{\alphasnomzerovalue}{\result{$1.49^{+0.14}_{-0.12}$}}